# Sloshing dynamics of liquid tank with built-in buoys for wave energy harvesting

Chongwei Zhang[a,b], Zhenyu Ding[a], Lifen Chen[a,c], Dezhi Ning[a,b,*]

[a]*State Key Laboratory of Coastal and Offshore Engineering, Dalian University of Technology, Dalian 116024, China*
[b]*China Three Gorges Corporation, Beijing 100038, China*
[c]*The University of Western Australia, 35 Stirling Highway,6009 Perth. Australia*

**Abstract**

This paper proposes a novel design of liquid tank with built-in buoys for wave energy harvesting, named the 'sloshing wave energy converter (S-WEC)'. When the tank is oscillated by external loads (such as ocean waves), internal liquid sloshing is activated, and the mechanical energy of sloshing waves can be absorbed by the power take-off (PTO) system attached to these buoys. A fully-nonlinear numerical model is established based on the boundary element method for a systematic investigation on dynamic properties of the proposed S-WEC. A motion decoupling algorithm based on auxiliary functions is developed to solve the nonlinear interaction of sloshing waves and floating buoys in the tank. An artificial damping model is introduced to reflect viscous effects of the sloshing liquid. Physical experiments are carriedout on a scaled S-WEC model to validate the mathematical and numerical methodologies. Natural frequencies of the S-WEC system are first investigated through spectrum analyses on motion histories of the buoy and sloshing liquid. The viscous damping strength is identified through comparisons with experimental measurements. Effects of the PTO damping on power generation characteristics of S-WEC is further explored. An optimal PTO damping can be found for each excitation frequency, leading to the maximisation of both the power generation and conversion efficiency of the buoy. To determine a constant PTO damping for engineering design, a practical approach based on diagram analyses is proposed, where the averaged conversion efficiency can reach 70%. Effects of the buoy's geometry on power generation characteristics of the S-WEC are also investigated. The geometry factors including the draught-to-width ratio (DWR) and inclination bottom angle of the buoys are investigated. For cases under consideration, the conversion efficiency of the S-WEC can even reach over 90%. In engineering practice, the present design of S-WEC can be a promising technical solution of ocean wave energy harvesting, based on its comprehensive advantages on surviv- ability enhancement, metal corrosion or fouling organism inhibition, power generation stability and efficiency, and so on.

*Corresponding author
 Email address:* `dzning@dlut.edu.cn` (Dezhi Ning)

*Keywords:* liquid sloshing, wave energy conversion, ocean energy, floating breakwater, offshore renewable energy

## 1. Introduction

Balancing greenhouse-gas emissions in the second half of the century has become an urgent demand to the international community, according to the Intergovernmental Panel on Climate Changes Special Report Global Warming of 1.5C, released late 2018 [1]. In response, formal net-zero emission targets have been announced by China, the European Union, the United

States, and technology firms such as Microsoft and European airports [2]. Transition from fossil fuels to renewable energy is becoming an important approach towards the goal of net zero. For coastal cities with large carbon dioxide emissions, the ocean is a great source of renewable energy.

Utilization of energy from ocean waves has attracted researchers' atten- tion for several decades [3]. To date, over one thousand patents on the novel design of wave energy converters (WECs) have been authorised, although a number of them may not be promising for further commercialisation in the near future [4]. The economic viability has become a key factor that determines the business potential of current WEC designs [5]. Compared to the wind energy, WEC projects targeting at harsh offshore environment have to face up to a particularly high capital expenditure (CapEx) and a long development process [6]. Many recent studies believe that integrating WECs with offshore structures (e.g. breakwaters and wind turbines), rather than deploying standalone WEC devices, can be an encouraging outlet for the leverage of construction and maintenance costs [7, 8].

For near-shore waters, the concept of integrating WECs with fixed break- waters have been accepted widely. For example, Kofoed [9] investigated single- or multi-level reservoirs installed on the top of a fixed breakwater, which endows the breakwater with a function of overtopping WECs. Vici-nanza et al. [7] and Contestabile et al. [10] introduced the reservoir in front of a rubble mound breakwater. Some practical designs have also embedded the oscillating water column (OWC) type WECs into the breakwater [11–16], while further technical improvements are still ongoing through laboratory in- vestigations [17–19] and numerical simulations [20, 21]. The integration of oscillating-buoy (OB) type WECs with a breakwater is another important research category. Mavrakos et al. [22] considered an array of cylindrical WECs heaving in front of a vertical breakwater. Schay et al. [23] numerically studied the hydrodynamic performance of a heaving point absorber near a fixed vertical wall. Zhao et al. [24] experimentally tested the performance of an OB-WEC array attached to a fixed pontoon. Zhang and Ning [25] numerically investigated the hydrodynamic performance of OB-WECs integrated with a specially designed breakwater. The breakwater has parabolic openings to focus the energy of incident ocean waves. Zhang et al. [26] analysed the hydrodynamic performance of an asymmetric heaving WEC deployed in front of a fixed breakwater.

When it comes to offshore sites in deep water, construction of fixed structures is often uneconomic. Some recent studies tend to integrate WECs with offshore floating structures. For example, Michailides and Angelides [27] applied a flexible floating breakwater for both shore protection and wave energy production. He et al. [28] introduced a box-type floating breakwa- ter with OWC chambers. Elhanafi et al. [29, 30] experimentally studied an offshore floating OWC device with single- or dual-chamber systems, respectively. Ning et al. [31] and Zhao and Ning [32] investigated pile-restrained floating breakwater with power take-off (PTO) devices. Madhi et al. [33] proposed the 'Berkeley Wedge' as an asymmetric WEC-breakwater. The Berkeley Wedge consists of an asymmetric floater with a particular shape, a PTO system, and a support structure to limit the floater's motion to heave only. Zhang et al. [34] numerically analysed the hydrodynamic performance of floating breakwaters (whose heaving motion was also used to absorb wave energy) with different bottom shapes, and concluded that, by carefully selecting parameters of the floater, it was possible to optimize wave energy conversion and attenuation performance of the floating breakwater.

At present, integrating WECs with offshore floating structures still has some tough challenges to break through before achieving any possible commercialisation. The fundamental challenge is associated with the survivability. Most WEC designs on a floating platform have to expose their sensitive moving parts in the harsh ocean environment. Long-time operation of such WEC systems can be severely threatened by not only extreme wave loads, but also chemical corrosion and fouling of marine organisms. The probability of frequent maintenances at distant offshore sites inevitably drives up the operational expenditure (OpEx) budget of an WEC project.

A possible idea to help increase WEC's survivability is to fully enclose the WEC system rather than to expose it in ocean waves. Towards this direction, some pioneer studies on various kinds of enclosed WECs have emerged in recent years. Fonseca and Pessoa [35] presented an asymmetric floater with an interior U-shaped liquid tank. An air turbine was installed at a duct connecting two OWC air chambers for electricity generation. Ribeiro e Silva et al. [36] further investigated the hydrodynamic optimization of the WEC with U-shaped interior oscillating water column. Crowley et al. [37] proposed a WEC design within a horizontal circular cylinder. The cylin-der had an annular tank half-filled with fluid. The pitching motion of sucha cylinder in waves can excited the liquid sloshing inside the tank, which further drives the air turbine in chambers above the free surface. Even so, systematic investigations of various internal WECs are still highly expected before reaching a converged technical solution.

Besides the above safety-related problems, some WEC designs also face the challenge of technical compromise when integrating with floating breakwaters. For example, the floating breakwater is expected to be as stationary as possible for a maximum reflection but minimum transmission of incident waves. Otherwise, an oscillating breakwater itself becomes a wave source that can generate waves into the shielded water area. However, the operation of some WEC systems relies on sound motions of the floating breakwater asa response to ocean waves. As a result, designers have to deal with two contradictory requirements on the floating breakwater, and make compromised technical decisions, which often inevitably misses the best energy harvesting condition [31]. Around this dilemma, the Berkeley Wedge can be considered as an effort towards increasing the breakwater's heave motion but avoiding severe wave transmissions [38]. Therefore, it is reasonable to study how to effectively harvest the wave energy when the motion amplitude of a floating breakwater has to be very small.

This study proposes a solution targeting at both challenges on survivability and technical compromise. The idea originates from the fundamental characteristic of liquid sloshing dynamics that even a tiny external excitation amplitude can trigger violent liquid motions near an infinite number of resonance conditions. Without sufficiently large damping, the sloshing energy can be restored within the liquid tank for a long time. If a buoy-type PTO system is deployed in the tank, the sloshing wave energy can be converted into electricity. Since the present device essentially extracts the mechani- cal energy from sloshing waves, we call this energy conversion unit 'sloshing wave energy converter (S-WEC)' for convenience. Fig. 1 (a) illustrates the cross-section sketch of an S-WEC. A complete S-WEC unit consists of two separate rooms. One is the 'dry' room (i.e. the energy conversion cabin) which contains machinery and electric devices for energy conversion, power conditioning, system control and so on. The other is the 'wet' room which is a fully-sealed liquid tank. The tank is half-filled with pure water or prepared solutions with certain density, and the air inside the tank is completely replaced with Nitrogen, Carbon Dioxide or inert gases. With these applications, metal corrosion induced by oxidation and fouling organism inside the tank can be avoided. The floating buoys in the liquid tank can slide along tank-side tracks, whose motion can drive the PTO system for electricity. The spring system linked to each buoy is used to limit the motion range of buoys and modulate natural frequencies of the PTO system.

The S-WEC is an isolated unit that can be installed on or within a floating structure. Fig. 1 (b) shows an application scenario where an array of S-WEC units is deployed inside a moored floating breakwater [39]. In operation, the floating breakwater is oscillated by external ocean waves, which subsequently activates the internal sloshing motion. After that, the sloshing waves further excite the floating buoys to drive the PTO system. Thus, the mechanical energy first transfers from ocean waves to the floating structure, and then to sloshing waves and oscillating buoys, before reaching the PTO system.

This study aims to conduct a systematic investigation on dynamic properties of the proposed S-WEC. In Section 2, the mathematical model of S-WEC is described, and a fully-nonlinear potential-flow solver based on boundary element method (BEM) is developed for numerical investigation. Section

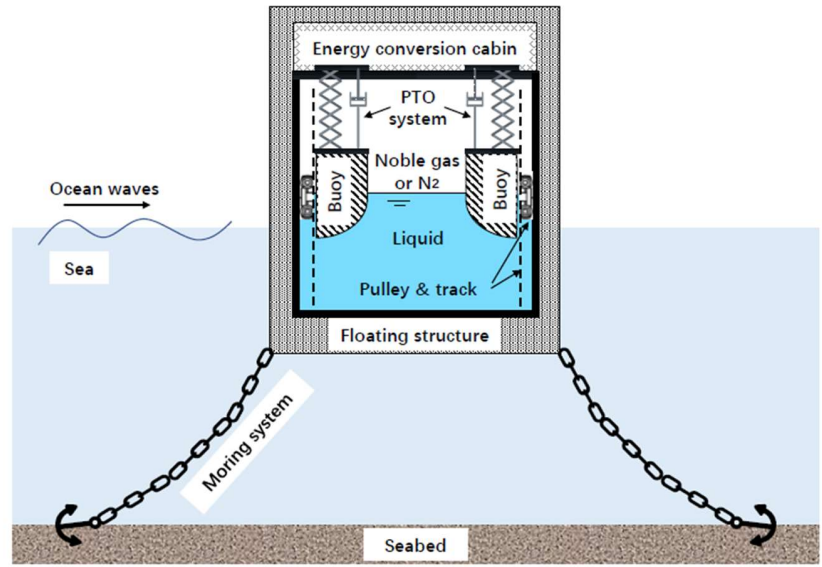

(a)

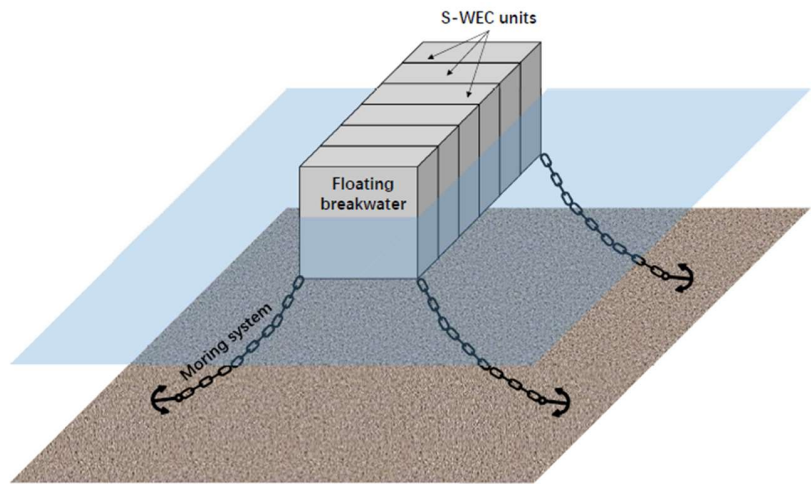

(b)

**Fig. 1.** (a) Internal sketch of S-WEC integrated on offshore floating structure; (b) an application scenario of S-WEC units on the floating breakwater [39]

3 introduces detailed set-ups of physical experiments carried out in the State Key Laboratory of Coastal and Offshore Engineering, Dalian University of Technology, for numerical validation. In Section 4, numerical tests are conducted to verify the present numerical model. Section 5 analyses natural frequencies and viscous damping calibration of the S-WEC model, as a system identification process. Section 6 carried out parametric study on S-WEC's performance, where effects of the PTO damping and geometry parameters of the buoys on the energy conversion performance are investigated. Section 7 further gives some discussions from the engineering practice point of view, to show additional advantages of the S-WEC design. Conclusions are drawn in Section 8.

## 2. Mathematical model

### 2.1. Problem description

This section first establishes a mathematical model for the proposed S-WEC, to help uncover mechanical properties of the system. The mechanical system associated with the sloshing tank can be simplified as Fig. 2. The tank is half-filled with water. Two identical buoys are floating on the water surface. The motion of each buoy is restrained by tracks on the internal tank wall, so that the buoy has only one degree of freedom (DOF), i.e. along the vertical direction. To distinguish these two buoys, the right buoy is marked 'Buoy-1' and the left 'Buoy-2'. Two buoys together with the tank wall are rigid, and the gap between the buoy and its adjacent tank wall is small enough to be omitted. The action of PTO system on the buoy is equivalent to a spring-damper system. For the situation when the mass of liquid tank is not comparable to that of the whole floating structure, the reaction of sloshing liquid to the structure can be omitted, so that the liquid tank canbe considered as undergoing a forced excitation.

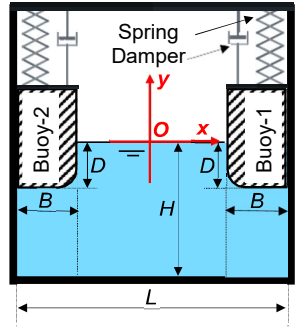

**Fig. 2.** Hydrodynamic model in the liquid tank of S-WEC and associated geometry parameters

Two right-handed Cartesian coordinate systems are introduced. One is the tank-fixed system $O-xy$, with its origin $O$ fixed at the centre of still water surface, $y$-axis pointing vertically upward and $x$-axis to the right. The other is the earth-fixed system $O_o - x_o y_o$, which coincides with $O-xy$ initially. When the tank is excited horizontally, the heave motion of Buoy-$i$ ($i =1$ or 2) is determined by Newton's Second Law as

$$M^{(i)} \dot{v}_2^{(i)}(t) = F_2^{(i)}(t) + F_{\text{damp}}^{(i)}(t) + F_{\text{PTO}}^{(i)}(t) - M^{(i)} g \qquad (1)$$

where the superscript $i$ indicates the variable associated with Buoy-$i$, $M$ represents the buoy mass, $F_2$ denotes vertical component of hydrodynamic force acting on the buoy, $F_{\text{damp}}$ is the drag force stemmed from vertex shedding around buoy corners, $F_{\text{PTO}}$ is the machinery force from the PTO system, $t$ is the time variable, $g$ is the gravitational acceleration, and $v_2$ and $v_2$ represent the vertical velocity and acceleration of the buoy, respectively. To be specific, the drag force increases quadratically with the relative flow velocity on the body surface, which is often non-negligible for high-speed flows.

Without loss of generality, the present S-WEC may apply a linear inductance generator to construct the PTO system of each buoy. The linear inductance generator can produce electrical current in the wire loop through relative motions between the wire loop and magnet. This can be achieved by letting a permanent magnet move through a coil or vice versa. According to McCormick [40], for the system when a conducting wire loop is situated between two heaving magnets, the produced electrical power can be obtained form

$$P^{(i)}(t) = \frac{N_e^2 B_e^2 l_e^2 v_2^{(i)}(t)}{R_e} v_2^{(i)}(t) \qquad (2)$$

where $N_e$ is the number of turns of the wire loop, $B_e$ is the magnetic induction, $l_e$ is the length of wire within the magnetic field, and $R_e$ is the load resistance in the wire loop. To avoid struggling with internal details of the generator, we use $C_{PTO}$ to represent the feature of the PTO system Sas

$$C_{PTO} = N_e^2 B_e^2 l_e^2 / R_e \qquad (3)$$

The generated electrical field can produce a motion resisting force of $C_{PTO} v_2^{(i)}(t)$ on each buoy in return. If springs are further used to restrain the motion of the buoys, the force acting on each buoy from the PTO system can be expressed as

$$F_{PTO}^{(i)}(t) = -C_{PTO} v_2^{(i)}(t) - K_{PTO} s_2^{(i)}(t) \qquad (4)$$

where $s_2$ is the vertical displacement of the buoy, and $K_{PTO}$ is the spring coefficient. Thus, the present PTO machinery behaves like a linear spring-damper vibration system. The first component on the right-hand-side of Eq. (4) is the resistive force that is related to the energy extraction, and $C_{PTO}$ can be defined as the PTO damping coefficient. The time-averaged power is obtained from

$$\bar{P}^{(i)} = \frac{1}{nT} \int_0^{nT} P^{(i)}(t) \mathrm{d}t \tag{5}$$

## 2.2. Nonlinear hydrodynamics

The interaction of sloshing waves and floating buoys in the tank is characterised by strong nonlinearity. The hydrodynamic force acting on each buoy can be calculated based on the fully-nonlinear potential-flow theory. By assuming the fluid to be inviscid, incompressible and flow-irrotational, the velocity potential $\phi$ whose gradient is the fluid velocity can be introduced. The hydrodynamic pressure $p$ in the fluid domain can be calculated according to Bernoulli's equation. Through an integration of the hydrodynamic pressure over the instantaneous wet surface of each buoy, the hydrodynamic force can be obtained as

$$F_2^{(i)} = -\rho \int_{S^{(i)}} \left( \frac{\partial \varphi}{\partial t} + \frac{1}{2} \nabla \phi \cdot \nabla \phi + gz \right) n_2 \mathrm{dS} \tag{6}$$

where $\Delta$ is the gradient operator, $n_2$ the second component of unit normal vector $\mathbf{n} = \{n_1, n_2\}$ and $\rho$ the water density. Note that $\mathbf{n}$ points outward from the fluid domain.

The velocity potential $\phi$ is determined by the following boundary value

problem in $O-xyz$

$$\nabla^2 \varphi = 0, \text{ in } V \tag{7}$$

$$\frac{\partial \eta}{\partial t} = v_1 \frac{\partial \eta}{\partial x} + \frac{\partial \varphi}{\partial y} - \frac{\partial \phi}{\partial x}\frac{\partial \eta}{\partial x}, \text{ on } S_F \tag{8}$$

$$\frac{\mathrm{d}\varphi}{\mathrm{d}t} = \frac{\partial \varphi}{\partial y}\frac{\partial \eta}{\partial t} + v_1 \frac{\partial \varphi}{\partial x} - \frac{1}{2}\nabla \phi \cdot \nabla \phi - g\eta - \mu\varphi, \text{ on } S_F \tag{9}$$

$$\frac{\partial \varphi}{\partial n} = v_1 n_1 + v_2^{(i)} n_2, \text{ on } S_B^{(i)} \tag{10}$$

$$\frac{\partial \varphi}{\partial n} = v_1 n_1, \text{ on } S_W \tag{11}$$

where the operator $\mathrm{d}/\mathrm{d}t = \partial/\partial t + (\partial \eta/\partial t)\,\partial/\partial y$ is a semi-Lagrangian expression, $\eta(x,t)$ denotes the free-surface elevation, $v_1$ is the horizontal velocity of tank surface and wet surface of Buoy-$i$, respectively. A Rayleigh viscosity term '$-\mu\phi$' is included in Eq. (9) to represent viscous dissipation effects of the liquid motion. The viscous dissipation effects in the present model are related to the boundary layer attached to internal tank surfaces and flow separations around buoy corners. When the wave motion starts from stationary, the initial conditions can be set as

$$\varphi = 0 \text{ and } \partial\varphi/\partial t = 0, \text{ in } V \text{ for } t \leq 0 \tag{12}$$

It should be noted that the $\phi_t$ (i.e. $\partial\phi/\partial t$) term in Eq. (6) cannot be explicitly obtained from the above boundary value problem. Conventional solution of $\phi_t$ through a backward finite difference method can cause sawtooth instabilities [41, 42]. To avoid such numerical instabilities, the value of $\phi_t$ can be solved based on the following boundary value problem

$$\nabla^2 \varphi_t = 0, \text{ in } V \tag{13}$$

$$\varphi_t = -\frac{1}{2}\nabla\varphi \cdot \nabla\varphi - g\eta - \mu\varphi, \text{ on } S_\text{F} \tag{14}$$

$$\frac{\partial \varphi_t}{\partial n} = \underbrace{\dot{v}_2^{(i)} n_2 + \dot{v}_1 n_1 - v_1 \frac{\partial^2 \varphi}{\partial n \partial x} - v_2^{(i)} \frac{\partial^2 \varphi}{\partial n \partial y}}_{R^{(i)}}, \text{ on } S_\text{B}^{(i)} \tag{15}$$

$$\frac{\partial \varphi_t}{\partial n} = \underbrace{\dot{v}_1 n_1 - v_1 \frac{\partial^2 \varphi}{\partial n \partial x}}_{R^0}, \text{ on } S_\text{W} \tag{16}$$

The second-order derivatives can be expressed by tangential derivatives of the fluid velocity according to the following mathematical transformation

$$\frac{\partial^2 \varphi}{\partial n \partial x} = \left(\nabla \frac{\partial \varphi}{\partial x}\right) \cdot \mathbf{n} = \frac{\partial^2 \varphi}{\partial x^2} n_1 + \frac{\partial^2 \varphi}{\partial x \partial y} n_2 = -\frac{\partial^2 \varphi}{\partial y^2} n_1 + \frac{\partial^2 \varphi}{\partial x \partial y} n_2$$
$$= -\frac{\partial^2 \varphi}{\partial y^2} \tau_2 - \frac{\partial^2 \varphi}{\partial x \partial y} \tau_1 = -\left(\nabla \frac{\partial \varphi}{\partial y}\right) \cdot \boldsymbol{\tau} = -\frac{\partial^2 \varphi}{\partial \tau \partial y} \tag{17}$$

$$\frac{\partial^2 \varphi}{\partial n \partial y} = \left(\nabla \frac{\partial \varphi}{\partial y}\right) \cdot \mathbf{n} = \frac{\partial^2 \varphi}{\partial y^2} n_y + \frac{\partial^2 \varphi}{\partial x \partial y} n_x = -\frac{\partial^2 \varphi}{\partial x^2} n_2 + \frac{\partial^2 \varphi}{\partial x \partial y} n_1$$
$$= \frac{\partial^2 \varphi}{\partial x^2} \tau_1 + \frac{\partial^2 \varphi}{\partial x \partial y} \tau_2 = \left(\nabla \frac{\partial \varphi}{\partial x}\right) \cdot \boldsymbol{\tau} = \frac{\partial^2 \varphi}{\partial \tau \partial x} \tag{18}$$

*2.3. Motion decoupling algorithm*

According to Eq. (1), the motion of floating buoys can be calculated based on the knowledge of hydrodynamic forces acting on them. To obtain the hydrodynamic force, the value distribution of $\phi_t$ over the wet buoy sur- faces must be known. Meanwhile, to solve the boundary value problem of $\phi_t$, the boundary condition in Eq. (15) is determined by the buoys' accelerations in return. Thus, the motion of two floating buoys and hydrodynamic forces acting on them are fully-coupled through $\phi_t$. In order to break this correlation loop, the method of auxiliary function method [42] is adopted here.

Boundary value problems of two auxiliary functions $\chi^{(i)}$ are first constructed as follows

$$\nabla^2 \chi^{(i)} = 0, \text{in } V \tag{19}$$

$$\chi^{(i)} = 0, \text{on } S_F \tag{20}$$

$$\frac{\partial \chi^{(1)}}{\partial n} = n_2, \text{on } S_B^{(1)} \tag{21}$$

$$\frac{\partial \chi^{(1)}}{\partial n} = 0, \text{on } S_B^{(2)} \text{and } S_W \tag{22}$$

$$\frac{\partial \chi^{(2)}}{\partial n} = n_2, \text{on } S_B^{(2)} \tag{23}$$

$$\frac{\partial \chi^{(2)}}{\partial n} = 0, \text{on } S_B^{(1)} \text{and } S_W \tag{24}$$

The following relationship is guaranteed according to Green's second identity

$$\int_{S_F + S_B^{(1)} + S_B^{(2)} + S_W} \left( \varphi_t \frac{\partial \chi^{(i)}}{\partial n} - \chi^{(i)} \frac{\partial \varphi_t}{\partial n} \right) dS = \int_V \left( \varphi_t \nabla^2 \chi^{(i)} - \chi^{(i)} \nabla^2 \varphi_t \right) dV = 0 \tag{25}$$

Substituting boundary conditions of $\chi^i$ and $\phi$ into Eq. (25) leads to

$$\int_{S_B^{(i)}} \varphi_t n_2 dS = \dot{v}_2^{(1)} \left( \int_{S_B^{(1)}} \chi^{(i)} n_2 dS \right) + \dot{v}_2^{(2)} \left( \int_{S_B^{(2)}} \chi^{(i)} n_2 dS \right)$$

$$\underbrace{- \left[ \int_{S_F} \varphi_t \frac{\partial \chi^{(i)}}{\partial n} dS \right] + \left[ \int_{S_B^{(1)}} \chi^{(i)} R^{(1)} dS \right] + \left[ \int_{S_B^{(2)}} \chi^{(i)} R^{(2)} dS \right] + \left[ \int_{S_W} \chi^{(i)} R^0 dS \right]}_{\mathfrak{R}^{(i)}}$$

$$= \dot{v}_2^{(1)} \left( \int_{S_B^{(1)}} \chi^{(i)} n_2 dS \right) + \dot{v}_2^{(2)} \left( \int_{S_B^{(2)}} \chi^{(i)} n_2 dS \right) + \mathfrak{R}^{(i)} \tag{26}$$

Then, the motion equations of two buoys can be rewritten as

$$M^{(i)}\dot{v}_2^{(i)}(t) = F_2^{(i)}(t) + F_{damp}^{(i)}(t) + F_{PTO}^{(i)}(t)$$

$$= -\rho \int_{S^{(i)}} \varphi_t n_2 dS - \rho \int_{S^{(i)}} \left(\frac{1}{2}\nabla\phi \cdot \nabla\phi + gz\right) n_2 dS + F_{damp}^{(i)}(t) + F_{PTO}^{(i)}(t)$$

$$= -\rho\dot{v}_2^{(1)}\left(\int_{S_B^{(1)}} \chi^{(i)} n_2 dS\right) - \rho\dot{v}_2^{(2)}\left(\int_{S_B^{(2)}} \chi^{(i)} n_2 dS\right) - \rho\mathfrak{R}^{(i)}$$

$$- \rho \int_{S^{(i)}} \left(\frac{1}{2}\nabla\varphi \cdot \nabla\varphi + gz\right) n_2 dS + F_{damp}^{(i)}(t) + F_{PTO}^{(i)}(t) \quad (27)$$

The instantaneous accelerations can be explicitly solved through the following linear system of equations

$$M^{(i)}\dot{v}_2^{(i)}(t) + \rho\dot{v}_2^{(1)}\left(\int_{S_B^{(1)}} \chi^{(i)} n_2 dS\right) + \rho\dot{v}_2^{(2)}\left(\int_{S_B^{(2)}} \chi^{(i)} n_2 dS\right)$$

$$= -\rho\mathfrak{R}^{(i)} - \rho \int_{S^{(i)}} \left(\frac{1}{2}\nabla\varphi \cdot \nabla\varphi + gz\right) n_2 dS + F_{damp}^{(i)}(t) + F_{PTO}^{(i)}(t) \quad (28)$$

*2.4. Numerical methodology*

Boundary value problems of $\phi$ and $\chi$ can be solved numerically by BEM. Take $\phi$ for example. Laplace's equation of Eq. (7) is first transformed into a boundary integral equation (BIE) over the entire fluid boundary $S$ based on Green's third identity as

$$\alpha(\boldsymbol{x}_p)\varphi(\boldsymbol{x}_p) = \int_S \left[G(\boldsymbol{x}_p, \boldsymbol{x})\frac{\partial \varphi}{\partial n} - \varphi(\boldsymbol{x})\frac{\partial G(\boldsymbol{x}_p, \boldsymbol{x})}{\partial n}\right] dS \quad (29)$$

where $\boldsymbol{x}_p = x_p, y_p$ denotes the position vector of a collocation point P on $S$, $\boldsymbol{x} = (x, y)$ represents the surface point on $S$, $\alpha(\boldsymbol{x}_P)$ is the solid angle and $G$ denotes the Green's function

$$G(\boldsymbol{x}_p, \boldsymbol{x}) = \ln\sqrt{(x - x_p)^2 + (y - y_p)^2} \quad (30)$$

To calculate the boundary integrations in Eq. (29), the boundary S is discretised into non-overlapping segment elements. A parametric coordinate system $O_\xi - \xi$ is defined along each element, with $O_\xi$ located at the middle point of the element and $\xi$ ranging from -1 to 1 between two element nodes. The value of **x**, $\phi$ or $\partial\phi/\partial n$ varies linearly along each element as

$$x(\xi) = \sum_{k=1}^{K} N_k(\xi) x_k \quad (31)$$

$$\varphi(\xi) = \sum_{k=1}^{K} N_k(\xi) \varphi_k \quad (32)$$

$$\frac{\partial \varphi(\xi)}{\partial n} = \sum_{k=1}^{K} N_k(\xi) \left(\frac{\partial \varphi}{\partial n}\right)_k \quad (33)$$

where $N_1(\xi) = (1-\xi)/2$ and $N_2(\xi) = (1+\xi)/2$ are shape functions, $K = 2$, and the subscript $k$ refers to the $k$th node of the element.

Locating $x_P$ at each boundary node yields the descritised form of Eq. (29) as

$$-\alpha_i \varphi_i + \sum_{e=1}^{N_e} \sum_{k=1}^{K} \int_{-1}^{1} \varphi_{d(e,k)} \left[ \frac{\partial G(\mathrm{x}(\xi), \mathrm{x}_i)}{\partial n} N_k(\xi) J_e(\xi) \right] d\xi$$
$$= \sum_{e=1}^{N_e} \sum_{k=1}^{K} \int_{-1}^{1} \left(\frac{\partial \varphi}{\partial n}\right)_{d(e,k)} [G(\mathrm{x}(\xi), \mathrm{x}_i) N_k(\xi) J_e(\xi)] d\xi,$$
$$\text{for } i = 1, 2, \cdots, N_{\text{nod}} \quad (34)$$

where $N_e$ and $N_{nod}$ denote the total number of elements and nodes of the boundary, respectively, the subscript $e$ and $i$ refer to variables of the $e$th element and $i$th node from the global counting system, respectively, and $d\,e,k$ denotes the global index of the $k$th node on the $e$th element. Here, $|J(\xi)|$ is the Jacobian determinant defined for the transform of coordinate

systems

$$|J(\xi)| = \sqrt{\left(\frac{dx}{d\xi}\right)^2 + \left(\frac{dy}{d\xi}\right)^2} \quad (35)$$

with

$$\frac{dx}{d\xi} = \sum_{k=1}^{K} \frac{dN_k(\xi)}{d\xi} x_k \text{ and } \frac{dy}{d\xi} = \sum_{k=1}^{K} \frac{dN_k(\xi)}{d\xi} y_k \quad (36)$$

Eq. (34) can be written in a concise form

$$\sum_{j=1}^{N_{nod}} H_{ij}\phi_j = \sum_{j=1}^{N_{nod}} G_{ij}\left(\frac{\partial \phi}{\partial n}\right)_j \quad (37)$$

with

$$G_{ij} = \delta_{j,d(e,k)} \left( \sum_{e=1}^{N_e} \sum_{k=1}^{K} G_i^{(e,k)} \right) \quad (38)$$

$$H_{ij} = \delta_{j,d(e,k)} \left( \sum_{e=1}^{N_e} \sum_{k=1}^{K} H_i^{(e,k)} \right) - \delta_{j,i} C_i \quad (39)$$

$$G_i^{(e,k)} = \int_{-1}^{1} \left[ G(\mathbf{x}(\xi), \mathbf{x}_i) N_k(\xi) J_e(\xi) \right] d\xi \quad (40)$$

$$H_i^{(e,k)} = \int_{-1}^{1} \left[ \frac{\partial G(\mathbf{x}(\xi), \mathbf{x}_i)}{\partial n} N_k(\xi) J_e(\xi) \right] d\xi \quad (41)$$

where $\delta_{j,d\,e,k}$ is Kronecker delta function. For $i \neq j$, both integrations of $G_i^{(e,k)}$ and $H_i^{(e,k)}$ are regular which can be calculated through standard Gauss quadrature. For $i = j$, generalized Gaussian quadratures for integrals with logarithmic singularity can be used to obtain $G_i^{(e,k)}$.

    Moving all known terms of Eq. (37) to the right hand side of each equation and unknown terms to the left, a linear system of equations can be solved, after which the velocity potential $\phi$ and its normal derivative $\partial\phi/\partial n$ at all boundary nodes of the fluid domain are known. The tangent derivative of $\phi$ on the boundary is calculated through a derivative of the cubic-spline

interpolation function with respect to *ϕ* over all free-surface nodes. With the tangent and normal derivatives of the velocity potential, the fluid velocity on the boundary is known. Thus, all right-hand-side terms in boundary conditions Eqs. (8) and (9) are known. A fourth-order Runge-Kutta method can be used to update *η* and *ϕ* to the next time step in the time domain. In order to avoid free-surface distortions due to numerical error accumulations, a standard five-point smooth procedure is also used to rearrange the free-surface nodes and smooth variables on them for every a few steps.

## 3. Experimental and numerical set-ups

Physical experiments are carried out on a scaled S-WEC model to validate the mathematical and numerical methodology in Section 2. The model is based on a box-shaped water tank as in Fig. 3 (a). The tank is made of transparent plexiglass plates of thickness 0.01m and density $1.18 \times 10^3 \text{kg/m}^3$, which can be considered rigid. The internal volume of the tank has length 0.5m, width 0.2m and height 0.6 m. Two U-shape aluminium tracks with a 0.025m×0.025m cross section are embedded in side walls of the tank. Two cuboid buoys with length $B = 0.128$m, width 0.16m and height 0.2m are manufactured, using plexiglass plates of thickness 0.003m. Both buoys are hollow, so that their draughts can be adjusted flexibly by filling plasticene inside. Two pulleys are installed on the side plate of each buoy and plugged into the track. Thus, the buoys can move freely along their track. Two Panasonic laser displacement sensors (LDSs) are installed at 0.2m above the tank top to record their vertical displacements. The LDS has a precision of 0.003m for targets in the range of 0.2~0.4m away from the sensor, and 0.08m-m for 0.4~0.6m. Then, the tank is filled with dyed water and fixed onto a shake table (i.e. Fig. 3 (b)) for ground tests in the State Key Laboratory of Coastal and Offshore Engineering, Dalian University of Technology. The shake table is a high-powered planar stage system driven by linear motors. It has a 0.5m×0.5m working area and 100kg loading capacity, and is capable to provide a horizontal excitation in two directions with an amplitude up to 20cm. During experiments, the water depth is set as $H$ =0.3m, and the draught of buoys is 0.06m. The shake table provides a harmonic excitation along the length side of the tank, with the displacement amplitude as $A$ =0.01m.

In order to compare with the experimental data, the tank length of the S-WEC model is fixed as $L = 0.5$m for all numerical simulations. Both buoys and the water in tank are stationary in the beginning. Unless specified, the initial depth of water in the tank is $H = 0.3$m. The horizontal displacement of the tank is governed by $s_0(t) = A\sin(\omega_0 t) = A\sin(2\pi f_0 t)$ with $A$, $\omega_0$ and $f_0$ as the amplitude, angular frequency, and frequency of the excitation, respectively. Hereafter, displacements of Buoy-1 and Buoy-2 are denoted by $s_1$ and $s_2$, respectively.

## 4. Numerical verification

Numerical convergence of the present BEM solver is first verified. Take the following case for example. Two rectangular buoys with breadth $B = 0.128$m and draught $D = 0.06$m are adopted. The damping coefficient is $\mu = 1.0 s^{-1}$, and $C_{PTO}$ and $K_{PTO}$ are ignored in this subsection. Excitation parameters of the tank are set as $A = 0.01$m and $f_0 = 0.8$Hz.

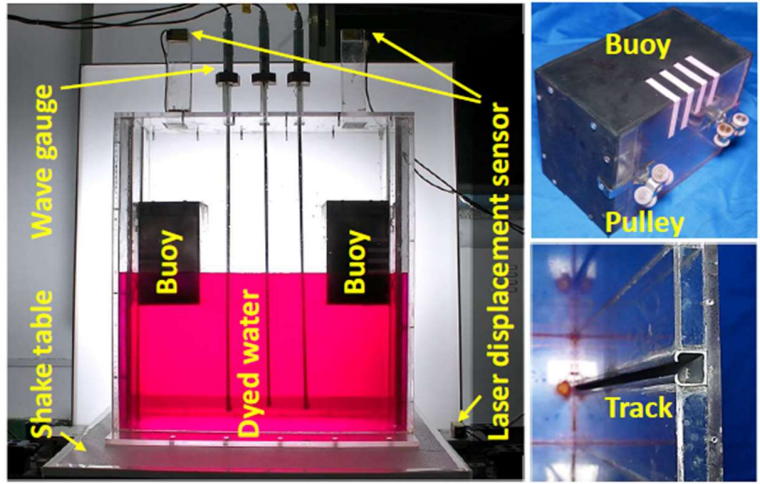

**(a)**

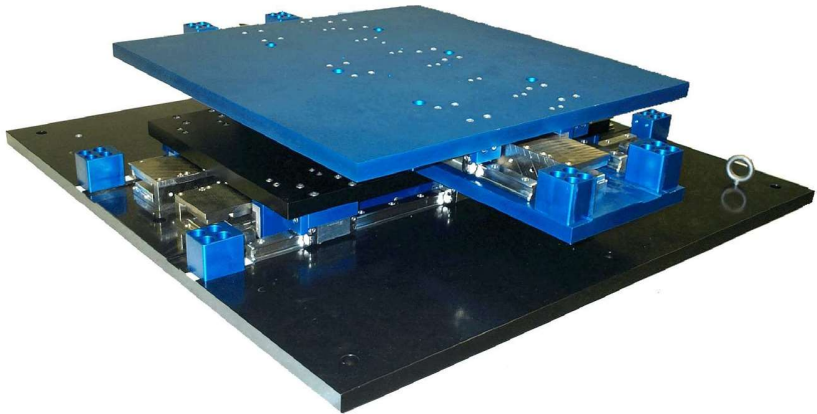

**(b)**

**Fig. 3.** (a) Arrangement of floating buoys and measuring sensors in liquid tank; (b) demonstration of the shake table

To verify the numerical convergence of spatial discretisation, the fluid boundary is discretised with three different element sizes, i.e. $\Delta l = L/12.5$, $L/25$ and $L/50$, respectively. Corresponding to each of these mesh sizes, the initial distribution of BEM nodes on the fluid boundary is depicted in Fig. 4. During the simulation, these BEM nodes are adjusted dynamically to maintain the element size. A sufficiently small time step $\Delta t = T/250$ is set for time marching, where $T$ denotes the period of excitation. Fig. 5 compares corresponding histories of Buoy-1's displacement and the free-surface elevation along Buoy-2. It shows that as the element is refined from $\Delta l = L/12.5$ to $L/25$, two displacement histories of the buoy have a good agreement with each other, but a visible distinction exists between two free-surface elevation histories. When the element size is further reduced by half, both histories of the buoy motion and free-surface elevation are not affected evidently. This means the numerical results can finally converge with the refinement of the boundary elements. For a spatial discretisation with the element size smaller than $\Delta l = L/25$, the numerical results are independent with the boundary mesh.

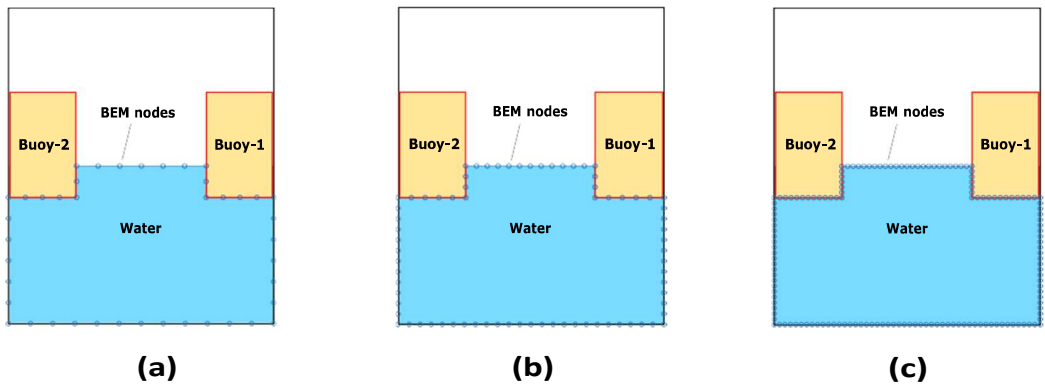

**Fig. 4.** Initial distribution of BEM nodes on fluid boundary, for $\Delta l =$ (a) $L/12.5$; (b) $L/25$; and (c) $L/50$

The numerical convergence of temporal discretisation is also tested by en-larging the time step from $\Delta t = T/250$ to $T/125$ and further $T/60$. The fluid boundary is discretised into sufficiently small elements (with $\Delta l = L/50$). Fig. 6 shows time histories of Buoy-1's displacement and the free-surface elevation along Buoy-2, corresponding to different time steps. It can be seen that results based on any of these three time steps do not have visible d- ifferences. Even so, to guarantee a precise capture of high-frequency wave components, the time step in the following studies is set as $\Delta t = T/125$.

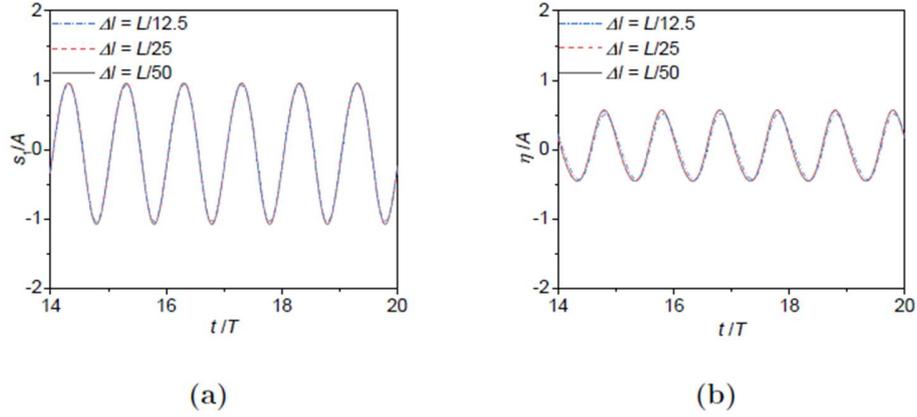

**Fig. 5.** Numerical convergence test on spatial discretisation, for time histories of (a) Buoy-1; and (b) free-surface elevation at $x = (B - L)$

The S-WEC is a novel model, whose hydrodynamic simulations cannot be found in open literature. To confirm the correctness of the present BEM solver, we can reproduce the same sloshing problem in Frandsen [43] through a removal of floating buoys from the water tank. Consider a rectangular liquid tank undergoing horizontal oscillations with the displacement $s_0(t) = A\cos(\omega_0 t)$. Parameters are set as $L = 1$m, $H = 0.5$m, $\omega_0 = 0.7\omega'_1$, $A = 0.0036g/\omega'^2$, where $\omega'_k = \sqrt{(k\pi g/L)\tanh(k\pi H/L)}$ corresponds to the $k$-th natural sloshing frequency. Fig. 7 compares wave elevation histories on the left tank wall obtained by the present numerical method and the fully-nonlinear finite differential method in the literature. Two results agree well with each other, which confirms the accuracy of the present numerical model for sloshing simulations.

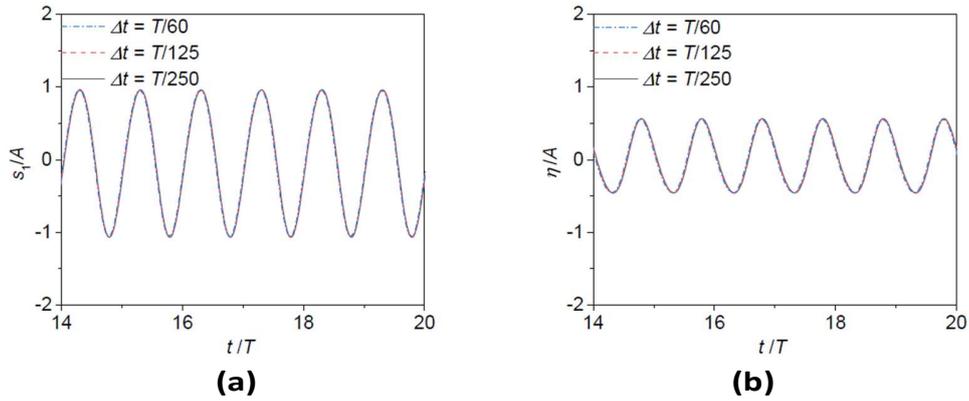

**Fig. 6.** Numerical convergence test on temporal discretisation, for time histories of (a) Buoy-1; and (b) free-surface elevation at $x = (B - L)$

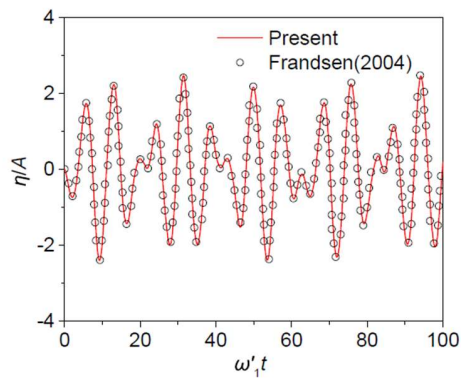

**Fig. 7.** Free-surface elevation histories on the left tank wall of a sloshing tank

## 5. System identification of S-WEC

### 5.1. Natural oscillation frequencies

The proposed S-WEC is essentially a mechanical vibration system with infinite DoFs. Mechanical properties on the natural frequencies should be identified in advance, before further analyses. To identify natural frequencies of the S-WEC system, all damping factors are first neglected. Frequency components in vibration histories of floating buoys and sloshing waves are analysed.

Take the case of buoys with $B = 0.128$m and $D = 0.06$m for example. It is expected that the dominant natural frequencies of the S-WEC are in the region around $\omega'_1$ of its sloshing tank. In order to recognise the exact natural frequencies, the S-WEC is excited harmonically with the amplitude $A = 0.01$m at three excitation frequencies $\omega_0 = 0.8\pi$ rad/s, $1.2\pi$ rad/s and $1.6\pi$ rad/s, respectively. Histories of two buoy's displacement and free-surface elevations at two locations are depicted in Fig. 8, corresponding to these three excitation frequencies. The free-surface elevations at $x = 0$ and $(B - L)$ are denoted by $\eta_0$ and $\eta_1$, respectively. All these oscillation histories have bounded amplitudes, indicating non-resonance situations. The general amplitudes of both the buoys and sloshing waves roughly increase with the excitation frequency, which suggests that these three frequencies are getting closer to the actual natural frequency of S-WEC.

The first thirty periods of histories in Fig. 8 are analysed through the fast Fourier transform (FFT). Fig. 9 (a) shows the spectra of Buoy-1's displacement corresponding to different excitation frequencies. Within the considered frequency range, two common frequency components at $\omega_1 = 0.92\omega'_1$ and $\omega_2 = 1.56\omega'_1$ can be observed, although the $\omega_1$ component has a much larger weight compared with the $\omega_2$ component. Fig. 9 (b) further compares the spectra of corresponding free-surface elevation histories, where the common frequency components at $\omega_1$ and $\omega_2$ also exist. As the excitation frequency approaches $\omega_1$ from the lower side, both weights of $\omega_1$ and $\omega_2$ increase evidently.

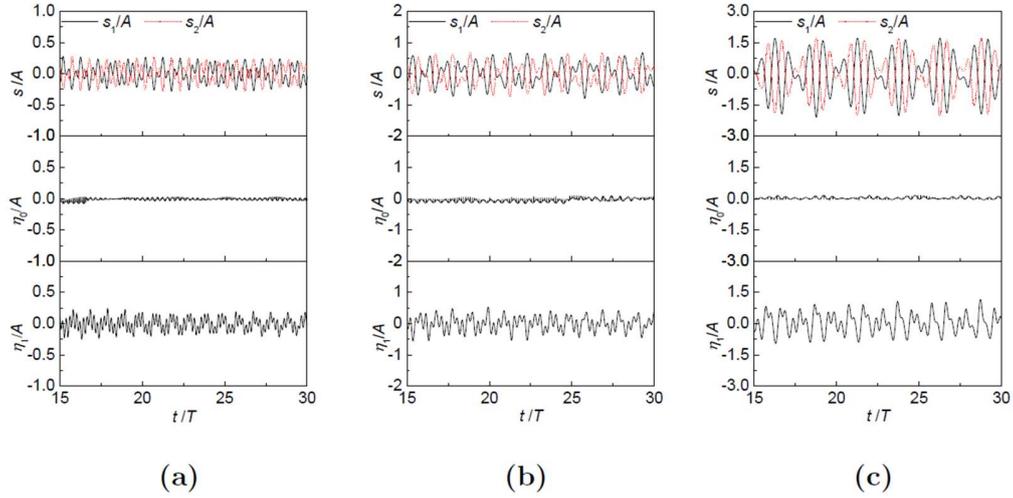

**Fig. 8.** Motion histories of two buoys and free-surface elevation at $x = 0$ and $(B - L)$, with $A = 0.01$m and (a) $\omega=0.8\pi$ rad/s; (b) $\omega=1.2\pi$ rad/s; (c) $\omega=1.6\pi$ rad/s

Since $\omega_1$ and $\omega_2$ are independent of the excitation, they can be expected as the first two natural frequencies of S-WEC. To confirm this speculation, the S-WEC is specially excited at $\omega_1$ and $\omega_2$, respectively. The excitation amplitude is decreased to an extremely small value $A = 0.0001$m, in order to avoid unnecessary nonlinear wave disturbances at this stage. For the excitation at $\omega = \omega_1$, the obtained motion histories of two buoys and free-surface elevation at $x = 0$ and $(B - L)$ can be found in Fig. 10 (a). It shows that these two buoys have anti-phase displacement histories, and the free-surface elevation amplitude at middle of the tank is nearly zero. This suggests that both the liquid and buoys in the tank are undergoing a nearly anti-asymmetric motion about the axis of symmetry. The amplitude of the buoy's displacement and wave run-up along Buoy-2 increase linearly with the time, as an evidence of resonance for a linear vibration system. Meanwhile, the increase ratio of the buoy' amplitude is much larger than that of the free surface, which suggests that the $\omega_1$ component tends to dominate the buoys' resonant motion. When the S-WEC is further excited at $\omega = \omega_2$, the obtained motion histories of two buoys and free-surface elevation at $x = 0$ and $(B - L)$ are given in Fig. 10 (b). The buoys' displacements are not harmonic with an increasing amplitude but with wrinkles on the curve, which indicates the occurrence of multiple frequency components. Even so, their amplitudes still show an increasing trend in general. On the free surface, the wave elevation at middle of the tank is still very tiny, while the wave run-up along Buoy-2 evidently increases with time in a linear sense. This means that the liquid is undergoing an antisymmetric resonance oscillation, the mode shape of which is similar to a clear tank without internal structures. Thus, it can be told that the $\omega_2$ component is more relevant to the liquid's sloshing resonance.

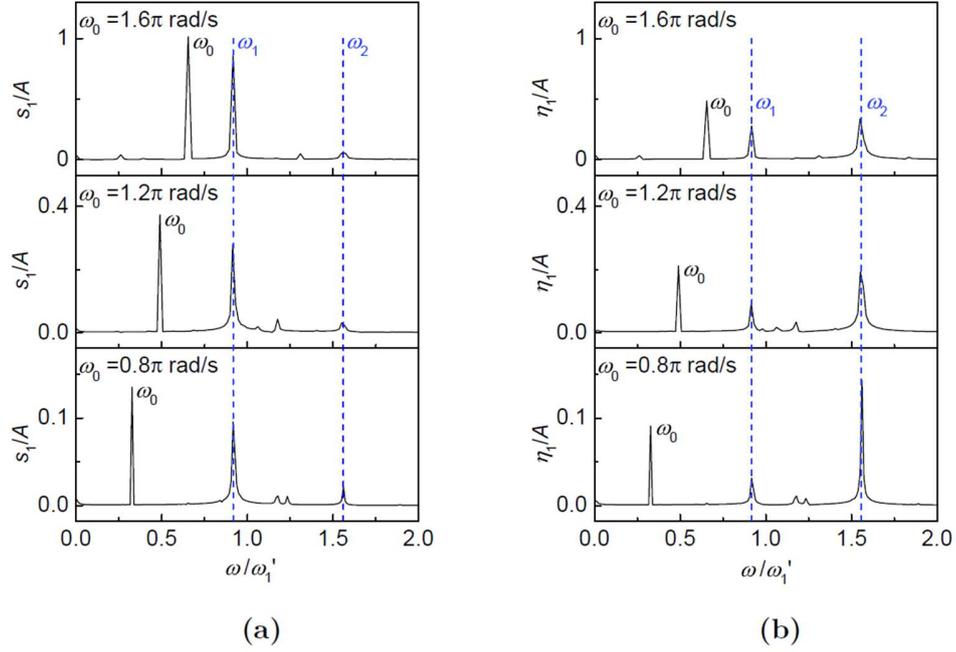

**Fig. 9.** FFT spectra of motions histories of (a) Buoy-1 and (b) free-surface elevation at $x = B-L$, corresponding to three excitation conditions at $\omega = 0.8\pi$, $1.2\pi$ and $1.6\pi$ rad/s

Fig. 11 further shows the corresponding FFT spectra of time histories in Fig. 10. It can be seen that when the tank is excited at $\omega = \omega_1$, only the $\omega_1$ frequency component is dominant for both the buoy and wave motions. Although the $\omega_2$ component can be also observed in the free-surface elevation history, it is too tiny to influence the spectrum. When the tank is excitedat $\omega = \omega_2$, the dominant frequency component for both the buoy and wave motions turn to $\omega_2$. For the buoy, the lower frequency $\omega_1$ is still evident. But for the liquid motion in term of $\eta_1$, the existence of $\omega_1$ component can be omitted.

From Figs. 10 and 11, it is reasonable to recognise $\omega_1$ and $\omega_2$ as the first two natural frequencies of the S-WEC system in a linear sense. Still, we should make a notation here that as time going on, the amplitude of either the

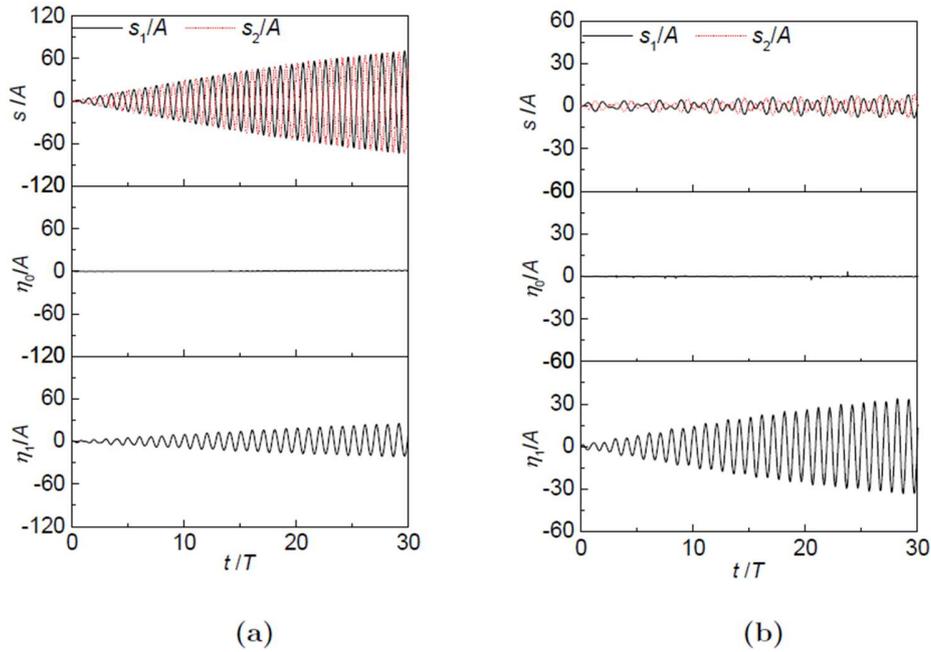

**Fig. 10.** Motion histories of two buoys and free-surface elevation at $x = 0$ and $(B - L)$, corresponding to $A = 0.0001$m and (a) $\omega = \omega_1$; (b) $\omega = \omega_2$

buoy's displacement or the free-surface elevation cannot grow to infinity by simply following the present linearly increasing trend. This can be attribute to different reasons. The practical reason is that the liquid tank has limited height, so that the floating buoys must be buffered by springs and latched for the safety before touching the tank ceiling or bottom. The numerical reason is that the BEM simulation must be terminated when the buoy exits the water at the extreme displacement or breaking wave occurs on the free surface. Even if without the halt of calculation, nonlinear effects can also be important. As the wave and motion amplitudes increase to a certain amount,

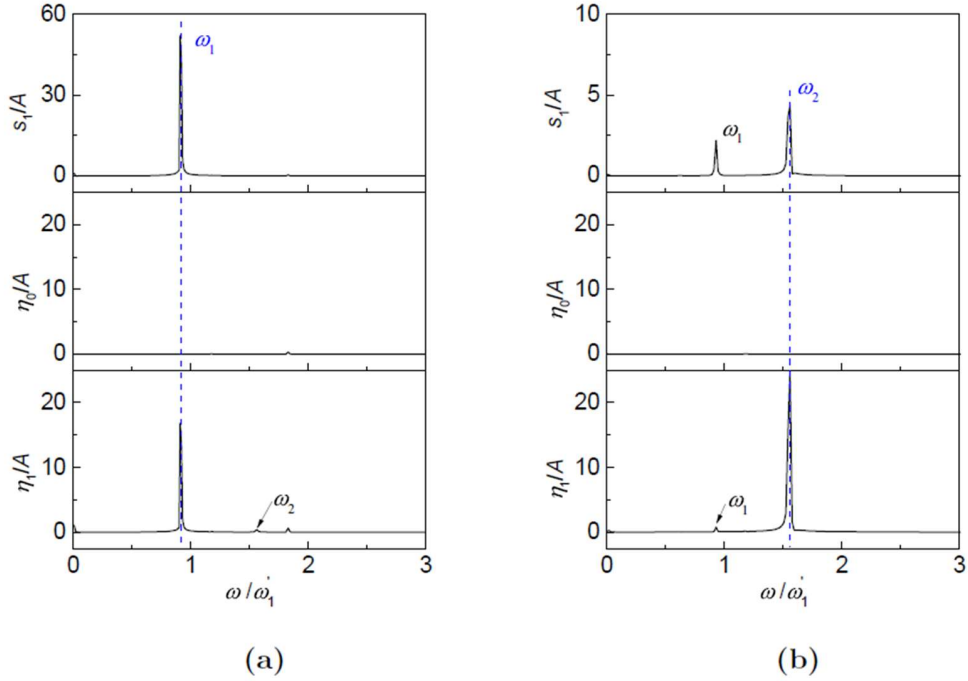

**Fig. 11.** FFT spectra of motion histories of two buoys and free-surface elevation at $x = 0$ and $(B - L)$, with $A = 0.0001$m and (a) $\omega = \omega_1$; (b) $\omega = \omega_2$

nonlinear mechanisms start to take effect by regulating natural frequencies. Governed by nonlinear mechanisms, the natural frequency $\omega_1$ or $\omega_2$ should drift away from the excitation frequency $\omega$, as the motion amplitudes further increase. As a result, the increasing trend of the buoys and wave motions has to slow down and reach the peak at some time, after which the motion amplitude will start to drop rapidly. Working principles of this nonlinear mechanism are similar to those described in Zhang[44]. However, even with the above concerns on complex nonlinear effects, it is still accurate enough to use $\omega = \omega_1$ and $\omega_2$ as the resonance conditions in the present concept proof stage for the S-WEC.

To include nonlinear effects into analysis, the excitation amplitude is increased up to $A = 0.01$m in subsequent cases. The results will show that

/

even though $A = 0.01$m is still a tiny amplitude compared with the tank size (i.e. $A = L/50$), it is sufficient to cause violent motions of both water waves and buoys. Fig. 12a shows motion histories of two buoys and free surface when the S-WEC tank is excited at $\omega = \omega_1$. It is clear that the heave amplitudes of two buoys increase rapidly with time. The buoys' motion amplitude can reach over ten times the excitation amplitude within four excitation periods, before the numerical simulation finally breaks down due to water exit of the buoy. Meanwhile, the water-wave motion is still very mild in this duration, which indicates that the sloshing wave energy is largely transferred to the buoys. This can be confirmed from snapshots in Fig. 13. Fig. 13 shows five snapshots of the S-WEC tank within a period from $t = t_0 = 2.58T$ to $t_0 + 4T/5$. For the energy harvesting purpose, the resonance mode corresponding to $\omega_1$ is more favourable. Because at this situation, even if the S-WEC is forced to oscillate with a tiny amplitude, the internal buoys can still undergo a large-amplitude motion, with a high potential of power generation. Meanwhile, the sloshing wave energy is largely transferred to the buoys, leaving aside a calm water domain.

Fig. 12 (b) is the case when S-WEC is excited at $\omega_2$. Compared to the observation in Fig. 12 (a), both the buoys and water waves have sound oscillations. The buoy's amplitude can reach over 5 times the excitation amplitude with an obvious envelope, while the free-surface elevation along the buoy can reach 7 times. The simulation breaks down after four and a half periods of excitation, at which the buoy exits the water. Fig. 14 shows five snapshots of the S-WEC tank within a from $t = t_0 = 3.10T$ to $t_0 +$ 4$T$/5. For both excitation situations in Figs. 12 (a) and 12 (b), twobuoys have nearly anti-asymmetric oscillations.

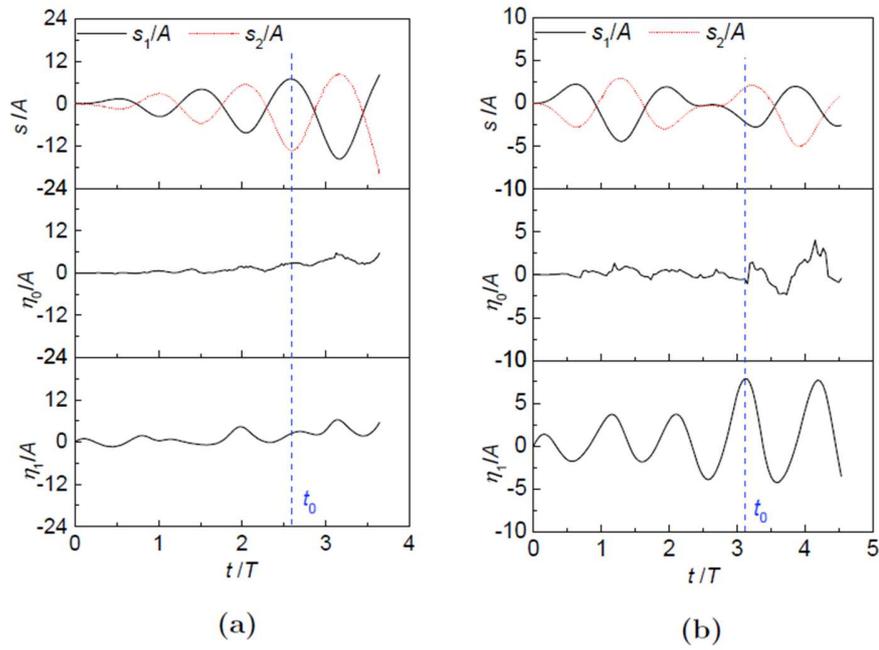

**Fig. 12.** Motion histories of two buoys and free-surface elevation at $x = 0$ and $(B - L)$, with $A = 0.01$m and $\omega =$ (a) $\omega_1$ and (b) $\omega_2$

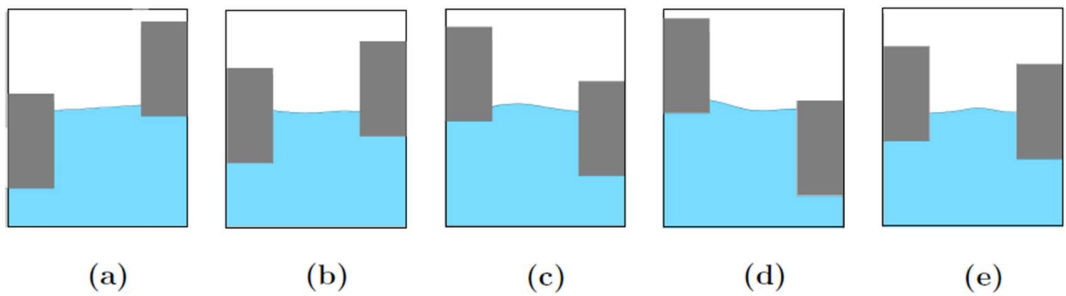

**Fig. 13.** Snapshots of S-WEC's liquid tank at (a) $t = t_0 = 2.58T$; (b) $t = t_0 + T/5$; (c) $t = t_0 + 2T/5$; (d) $t = t_0 + 3T/5$; and (e) $t = t_0 + 4T/5$, for $A = 0.01$m and $\omega = \omega_1$

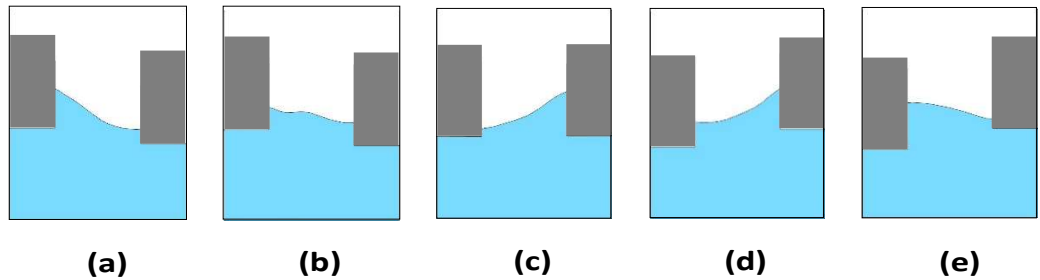

**(a)**        **(b)**        **(c)**        **(d)**        **(e)**

**Fig. 14.** Snapshots of S-WEC's liquid tank at (a) $t = t_0 = 3.10T$; (b) $t = t_0 + T/5$; (c) $t = t_0 + 2T/5$; (d) $t = t_0 + 3T/5$; and (e) $t = t_0 + 4T/5$, for $A = 0.01$m and $\omega = \omega_2$

## 5.2. Viscous damping calibration

Without the inclusion of viscous effects, transient sloshing effects cannotbe dissipated naturally in a closed liquid tank. This may result in a great distinction between the numerical prediction and physical reality. Take a non-breaking sloshing process for example. The numerical results cannot reach the steady state with constant wave amplitudes at or near resonance conditions, which disobeys the physical observation in reality. For the present potential-flow simulations, an artificial treatment is introduced to mimic the damping effects to the greatest extent. To be specific, the artificial damping term with an unknown coefficient $\mu$ (in Eq. (9)) is used to approximate the energy dissipation due to the fluid's viscosity. The value of $\mu$ can be distinctfor different scales of tank models, which should be calibrated through a comparison with experimental data.

Fig. 15 compares the calculated displacement amplitudes of Buoy-1 corresponding to three damping coefficients (i.e. $\mu$=1.0, 1.9 and 3.0 s$^{-1}$). The liquid tank is excited in the frequency range of $\omega \in [0.66\omega'_1, 1.39\omega'_1]$. The excitation amplitude is kept as $A = 0.01$m. Corresponding to each damping strength, as the excitation frequency increases, the motion amplitude increases to its peak and then drops gradually. For the convenience of subsequent discussions, we define the frequency range between $\omega = 0.82\omega'_1$ and $0.98\omega'_1$ the 'resonance region', where the amplitude peak appears. The terms of lower- and higher-frequency range are also used, corresponding to $\omega = [0.66\omega'_1, 0.82\omega'_1]$ and $\omega = [0.98\omega'_1, 1.39\omega'_1]$, respectively.

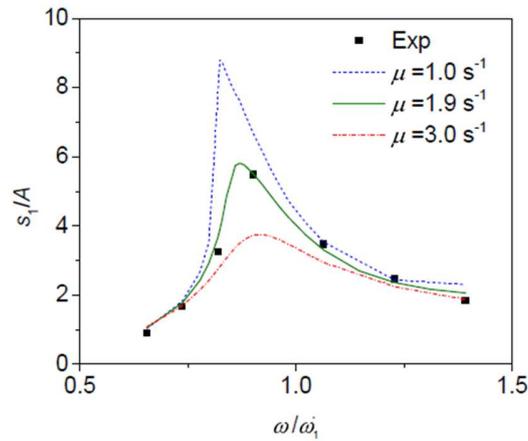

**Fig. 15.** Motion amplitude of Buoy-1 versus excitation frequency. Numerical results corresponding to $\mu$=1.0, 1.9 and 3.0 s$^{-1}$ are denoted by solid, dashed and dashed-dot lines, respectively. Experimental data are in dot.

In the lower-frequency region, the motion amplitude increases sharply and shows a 'jump' phenomenon towards the resonance condition. In the resonance region, the motion amplitude of the buoy decreases with the in crease of the damping coefficient. With the choice of μ =1.9−1, the calculated buoy's amplitudes have the best agreement with the experimental data, and the motion amplitude can reach about 6 times the excitation amplitude. It is also interesting to find that, corresponding to a larger damping strength, the peak of the motion amplitude tends to appear at a higher frequency, as an evidence of the non-linear soft-spring behaviour. In the higher-frequency region between $\omega = 0.98\omega'_1$ and $1.39\omega'_1$, the damping strength does not affect the buoys' motion evidently, and the motion amplitude is above 2 times the excitation amplitude.

With $\mu = 1.9$ s$^{-1}$, Fig. 16 compares snapshots of S-WEC's liquid tank with those from the experiment during one period of excitation (started from $t_0 = 46.6T$) for the excitation condition of $\omega = 0.9\omega'_1$, while Fig. 17 compares snapshots from $t_0 = 23.8T$ for the excitation condition of $\omega = 1.56\omega'_1$. In both figures, subfigures (a-1) to (a-5) are numerical results, and (b-1) to (b-5) are experimental pictures. Through comparison, it can be confirmed that the numerical solver can give a satisfactory prediction on liquid profiles and buoys' locations, with a properly defined $\mu$.

Still, it should be noted that two buoys in the physical experiment have limited dimensions along the width of the tank, which means the liquid's motion in the tank inevitably has three-dimensional disturbances. The present 2D mathematical model may not be able to perfectly reproduce physical real- ity from every aspect. Even so, with a properly selected damping coefficient, the present model has shown to be an effective tool on approximating the S-WEC's practical behaviour. In subsequent analyses, the damping coefficient is constantly set as $\mu = 1.9$ s$^{-1}$ to reproduce inevitable viscous damping

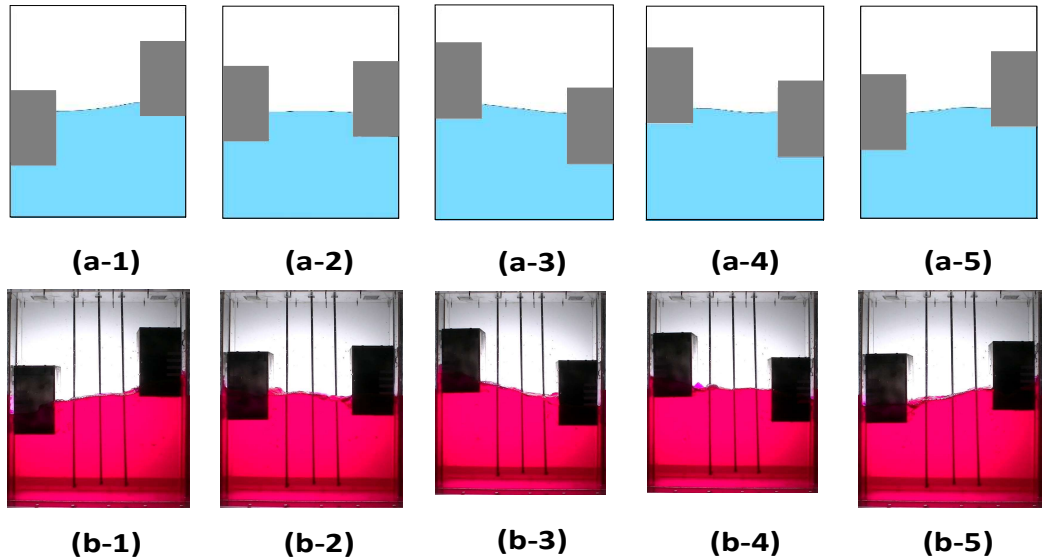

**Fig. 16.** Snapshots of S-WEC's liquid tank at (a) $t = t_0 = 46.6T$; (b) $t = t_0 + T/5$; (c) $t = t_0 + 2T/5$; (d) $t = t_0 + 3T/5$; and (e) $t = t_0 + 4T/5$, for $A = 0.01$m and $\omega = \omega_2$. Here, (a-1) to (a-5) are numerical results, and (b-1) to (b-5) are experimental photograph at the corresponding instants

effects in the S-WEC tank model.

## 6. Parametric study on S-WEC's performance

### 6.1. Optimised PTO damping

The conversion efficiency, defined as the ratio of the captured wave power to external power inputs, is an important indicator of energy conversion performance of an S-WEC design. For an S-WEC undergoing forced oscilla- tions, the work done by the external force is hard to evaluate. In this study,

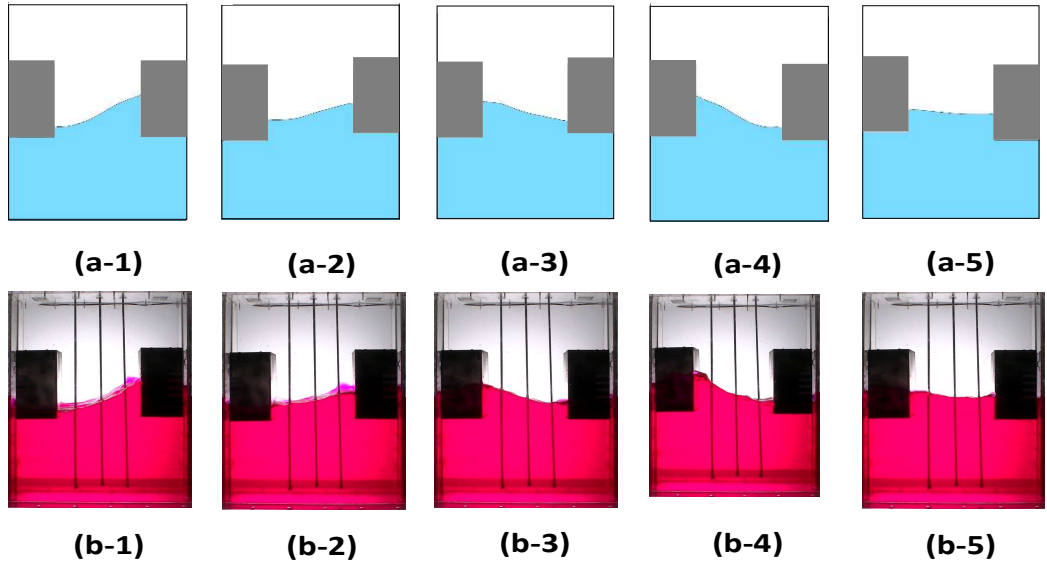

**Fig. 17.** Snapshots of S-WEC's liquid tank at (a) $t = t_0 = 23.8T$; (b) $t = t_0 + T/5$; (c) $t = t_0 + 2T/5$; (d) $t = t_0 + 3T/5$; and (e) $t = t_0 + 4T/5$, for $A = 0.01$m and $\omega = \omega_2$. Here, (a-1) to (a-5) are numerical results, and (b-1) to (b-5) are experimental photograph at the corresponding instants

the power input into the S-WEC system is estimated by reference power $P_I$

$$\hat{P}_I = m \int_0^{T/4} \dot{v}_1(t)\, v_1(t)\, \mathrm{d}t = m A^2 \omega_0^3 / \pi \qquad (42)$$

where $m$ represents the total mass of the liquid and buoys of the S-WEC. The reference power is obtained by assuming the liquid to be 'frozen'. With the reference power, the conversion efficiency can be defined in the form of $\eta_{ave} = P_{ave}/\hat{P}_I$, where $P_{ave}$ denotes the averaged power obtained by the PTO system during the steady stage. In the following studies, 50 periods of the excitation are considered to calculate $P_{ave}$. Fig. 18 is an example of time history of the captured power by Buoy-1, where the excitation amplitude is $A = 0.01$m, excitation frequency is $\omega_0 = 0.9\omega_1'$ and PTO damping is $C_{PTO} =30$kg/s. A dotted blue line is used to mark out the location of $P_{ave}$, and the shaded area indicates the total energy captured by the PTO system. It can be seen that the $P_{ave}$ is less than half of the maximum instantaneous power $P_{max}$.

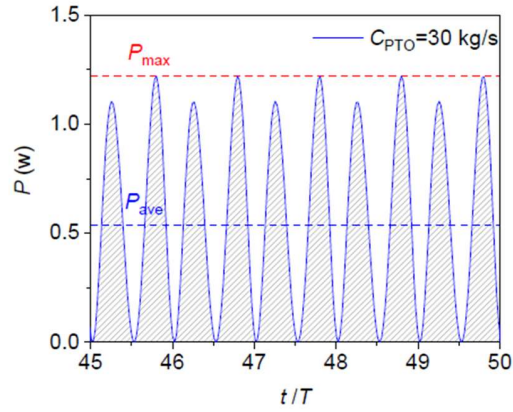

**Fig. 18.** Time histories of the power captured by the PTO system of Buoy-1, with $A = 0.01$m, $\omega_0 = 0.9\omega_1'$ and $C_{PTO} =30$kg/s

Since the power generation of S-WEC is directly determined by vertical oscillations of two buoys in the liquid tank, effects of the PTO damping on motion characteristics of the buoys are first explored here. Fig. 19 shows amplitude variations of Buoy-1's steady-state oscillations versus the excitation frequency, under different settings of PTO damping. With each PTO damping, the amplitude-frequency curve has an extreme in the considered frequency range $\omega \in [0.66\omega_1', 1.31\omega_1']$. The frequency where the peak amplitude occurs can be identified as the resonant frequency of the buoy. The extreme amplitude of the buoy decreases with the increase of $C_{PTO}$, which means a larger PTO damping can lead to a severer suppression effect on the motion of buoy. Meanwhile, the resonant frequency of each buoy tends to increase with the PTO damping, relevant to nonlinear sloshing properties of the liquid. Thus, variation of PTO damping can affect both the oscillation amplitude and resonance frequency of the buoy.

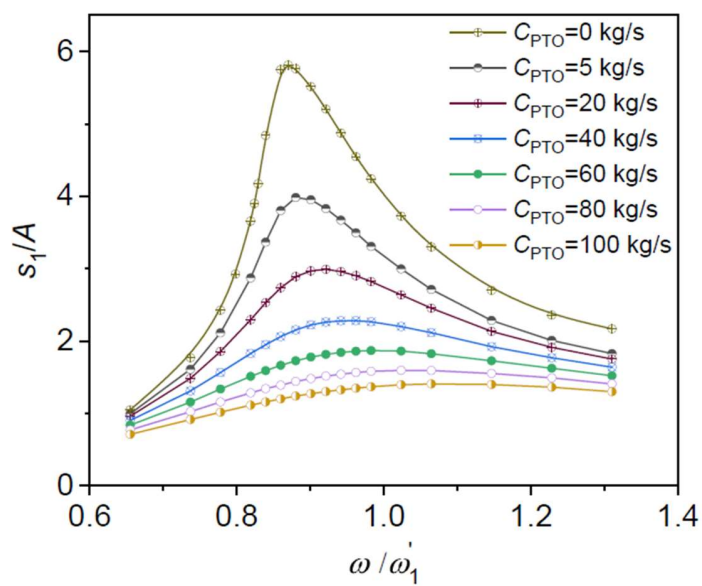

**Fig. 19.** Displacement amplitude of Buoy-1 against the excitation frequency under different values of PTO damping

Fig. 20 shows the averaged power generation of Buoy-1 against the P- TO damping coefficient $C_{PTO}$, under different excitation frequencies. Corresponding to each excitation frequency, as $C_{PTO}$ grows from 5 to 300kg/s, the power captured by Buoy-1 first increases towards a peak and then falls. The optimal PTO damping $C_{PTO-opt}$ can be identified when reaching the maximised power generation. Positions of these maximum power generations are linked with red arrows. It can be seen that as the excitation frequency varies from 0.66 to 1.31$\omega_1'$, and optimal PTO damping first decreases and then increases. The border values of $C_{PTO-up}$ and $C_{PTO-low}$ are also marked out in the figure to show the condition when 90% of the maximised power generation can be obtained.

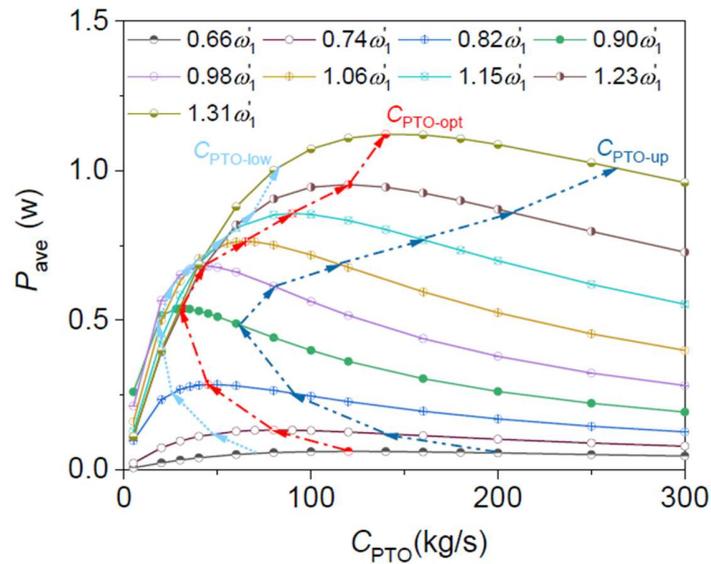

**Fig. 20.** Averaged power generation of S-WEC's Buoy-1 against the PTO damping

Fig. 21 further shows the conversion efficiencies corresponding to Fig. 20. For each excitation frequency, the optimal PTO damping can lead to the maximisation of both the power generation and conversion efficiency of the buoy, which is reasonable regarding to the fact that the conversion efficiency is obtained through a division of the captured power by the reference power. The reference power is constant for a specific excitation condition, so that the captured power and conversion efficiency have the same trend as the PTO damping increases. However, the captured power at the optimal PTO damping gradually increases with the excitation frequency, while the conversion efficiency first increases and then decreases with the frequency. This is because, the reference power on the denominator of the conversion efficiency increases with the excitation frequency. For the PTO damping in the range of [5kg/s, 40kg/s], the conversion efficiency of the buoy increases with the PTO damping. With $C_{PTO}$=40kg/s, Buoy-1 of the S-WEC system achieves the highest conversion efficiency around 35%.

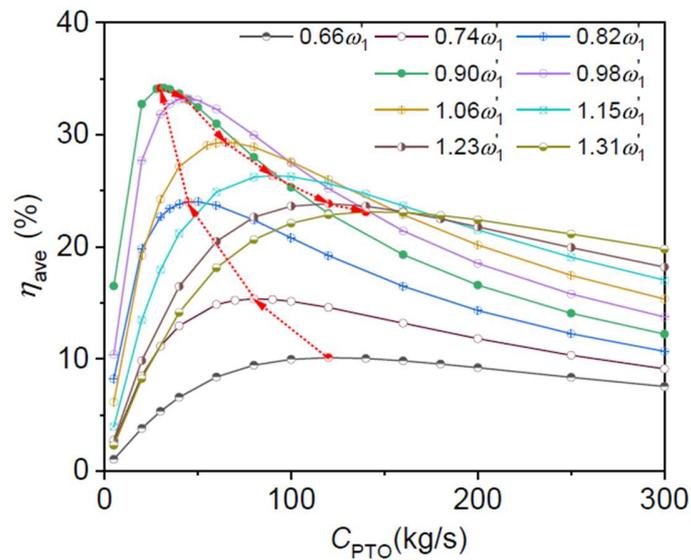

**Fig. 21.** Conversion efficiency of S-WEC's Buoy-1 against the PTO damping

Fig. 22 shows the conversion efficiency of Buoy-1 with different PTO damping and excitation frequencies from a different perspective. With a fixed PTO damping, there exists a maximum conversion efficiency within the considered frequency range. As the PTO damping increases from $C_{PTO}$ = 5 to 40kg/s, the conversion efficiency increases at all excitation frequencies, and the efficiency-frequency curve becomes more flat. A more flat efficiency-frequency curve suggests a less sensitive of the conversion efficiency to the excitation frequency. As the damping coefficient further increases, the maximum conversion efficiency decreases, but the flatness of the efficiency-frequency curve continuously grows. Similar to Fig. 20, the peaks of the efficiency-frequency curve can be linked in order with red arrows. It shows that as the PTO damping increase, the excitation frequency that corresponds to the maximum conversion efficiency increases.

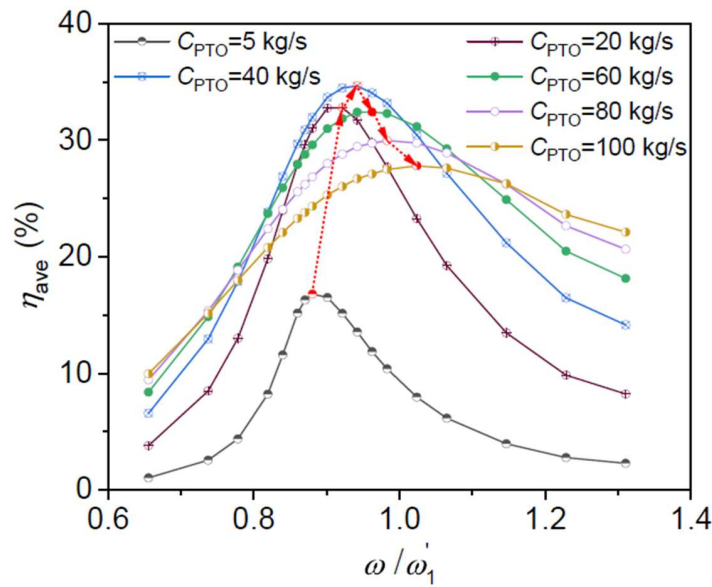

**Fig. 22.** Conversion efficiency of S-WEC's Buoy-1 against the excitation frequency under different values of PTO damping

As a classification of the above observations, Fig. 23 shows the variation of the optimised PTO damping $C_{PTO-opt}$ as a function of the excitation frequency, together with the maximum power capture and conversion efficiency of Buoy-1 under the optimised PTO damping setting. In the top sub-figure, the red line represents the optimal PTO damping for the S-WEC devicethat can lead to the maximum power generation at the corresponding ex-citation frequency. As the excitation frequency increases from 0.66 to 1.31 $\omega_1'$, the optimal PTO damping first decreases and then increases. The value of $C_{PTO-opt}$ reaches the minimum extreme at around $\omega = \omega_1$, where $\omega_1$ is the resonance frequency of the buoy in the liquid tank. The blue lines represent the PTO damping with which the captured power can reach 90% of the maximum power generation in an optimal condition with $C_{PTO-opt}$. The middle sub-figure shows the maximum $P_{max}$ and $P_{ave}$ against the excitation frequency, corresponding to the optimised PTO damping at each excitation frequency. Both the maximum $P_{max}$ and $P_{ave}$ increase with the excitation frequency. This is reasonable because the total power input to the S-WEC system also increases with the excitation frequency. In the bottom sub-figure, the maximum conversion efficiency of $\eta_{ave}$ for the S-WEC system with both buoys. With the increase of the excitation frequency, the conversion efficiency increases at first and then decreases. The highest conversion efficiency occurs in the excitation condition of $\omega = \omega_1$, and the maximum conversion efficiency can be 70% at most. Referring to Fig 19, it can be understood that the highest conversion efficiency of the device appears when the buoys are right under resonance oscillations. In other words, with an optimal PTO damping, the buoy of the S-WEC oscillates at resonance and its conversion efficiency also reaches the maximum.

Thus, different excitation frequencies correspond to different choices of the optimal PTO damping, respectively. If an invariable PTO damping must be determined for the practical engineering design, the following approach can be followed. Referring to the blue lines in the top sub-figure of Fig. 23, a shaded regime can be found in between. At any excitation frequency,

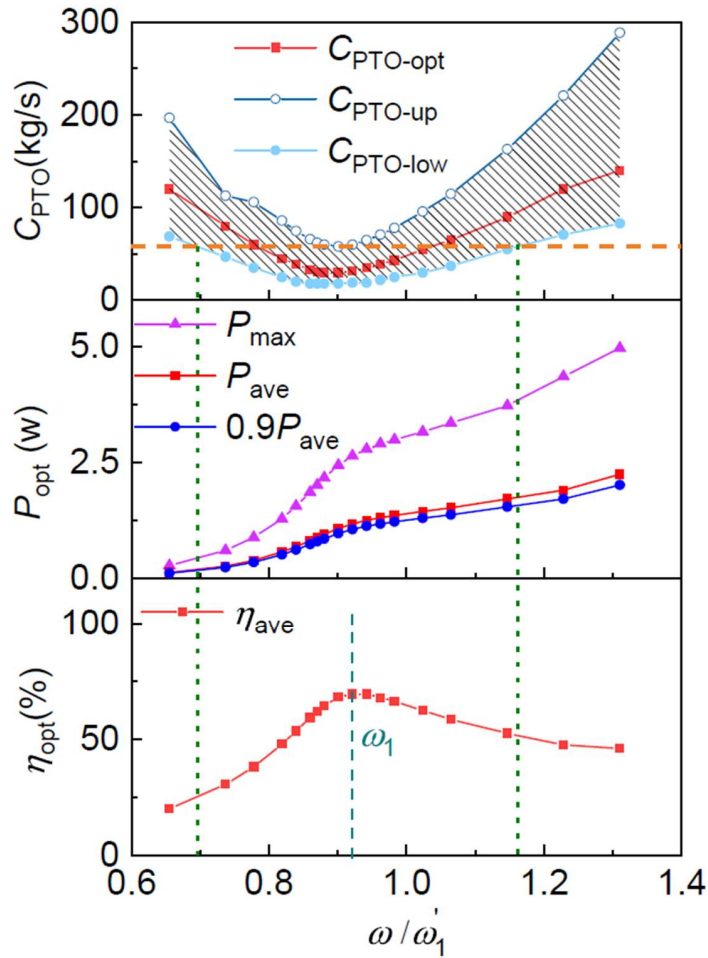

**Fig. 23.** Optimal PTO damping, maximum power capture and maximum conversion efficiency of S-WEC, corresponding to different excitation frequencies

as long as the PTO damping takes the value within the shaded regime, the captured power of S-WEC can reach over 90% of the maximum $P_{ave}$ in that excitation condition. Then, a horizontal line tangent to the regime border can be drawn in the shaded regime. With a fixed PTO damping along this line, the S-WEC can achieve at least 90% of the maximum $P_{ave}$ over the largest frequency range (i.e. the frequency range restrained by two green lines in the middle sub-figure). For the present case of S-WEC design, the PTO damping can be fixed as $C_{PTO} = 58$ kg/s, where the obtained $P_{ave}$ varies from 0.2W to 1.56W and the averaged conversion efficiency can reach 25% to 70% within the frequency range of $[0.7\omega', 1.16\omega']$. For even higher frequencies, the maximum conversion efficiency can be around 50%,

*6.2. Effect of buoy's draught-to-width ratio on power generation*

In previous experimental and numerical tests, the draught-to-width ratios (DWRs) of the buoys is fixed to be 0.5. This subsection will further consider buoys of various DWRs on power generation of the S-WEC system. All buoys in consideration have the constant displacement volume, but different DWRs (varying from $D/B = 0.22$ to 8). The tank size and liquid depth of the S-WEC are the same as those set in subsection 6.1. Referring to the method in subsection 5.1, natural frequencies of the S-WEC system are first tested. As shown in Fig. 24, $\omega_{s_1}$ and $\omega_{\eta_1}$ are used to denote the fundamental natural frequencies of the buoy and sloshing liquid, respectively. It can be seen that, as the DWR increases, both $\omega_{s_1}$ and $\omega_{\eta_1}$ gradually decrease. The fundamental natural sloshing frequency of a liquid tank with length $(L - 2B)$ and depth $H$, i.e. $\omega'_{\eta_1}$, is also show in the figure for comparison. It is interesting to find that $\omega'_{\eta_1}$ can be a good approximation of $\omega_{\eta_1}$, especially for large DWR cases. This can be understood that for larger draught buoys, the constrained water body between the buoy and tank bottom has less effects on natural sloshing frequencies in the tank.

The power generation of S-WECs with different DWRs of buoys is fur- ther considered. Physical parameters including the viscous damping coeffi-cient and excitation amplitude are the same as those set in subsection 6.1. According to findings in the previous subsection, the maximum conversion efficiency of the S-WEC system occurs when excited at the natural frequency of the buoy. In order to compare the largest power generation of different buoys, the S-WECs with different DWRs of buoys are excited at the corresponding natural frequencies according to Fig. 24. Then, the optimal PTO damping at the natural frequency of each buoy can be found, referring to Fig. 25, It shows that, from $D/B$=0.22 to 2.00, the optimal PTO damping of the buoy decreases rapidly. For buoys with even a larger DWR, values of the optimal PTO damping are almost constant. Using buoys with their optimal PTO damping, the captured power and conversion efficiency of the S-WEC are obtained as shown in Fig. 25. As DWR increases, both the maximum power generation and conversion efficiency first grow and then fall. The maximised power generation appears at $D/B$=0.89, with the maximum power of 2.8W, and averaged power of 1.2W. In terms of the conversion efficiency, the DWRs from 0.22 to 4.0 are all satisfactory, which can lead to the conversion efficiency of 50 to 75%. As DWR further grows over 4, both power capture and conversion efficiency decrease gradually, although the conversion efficiency is still above 30%. Thus, the buoy DWR of a S-WEC generally does not affect the maximum conversion efficiency severely.

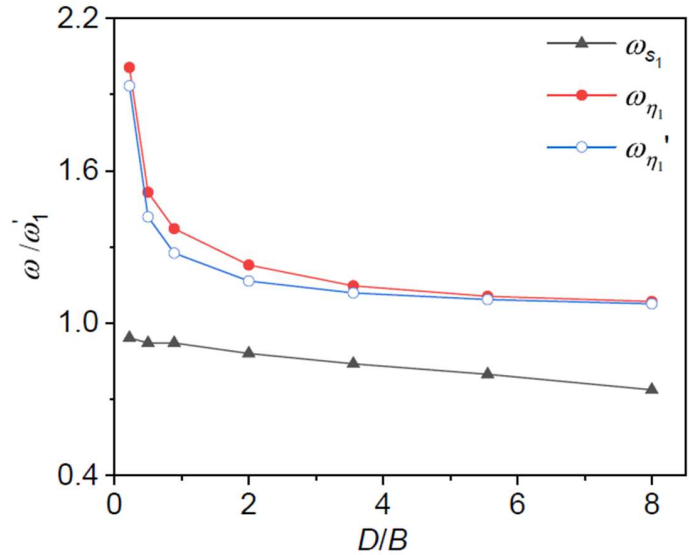

**Fig. 24.** Natural frequencies of the buoy and liquid in S-WECs with different draught-to-width ratios of buoys

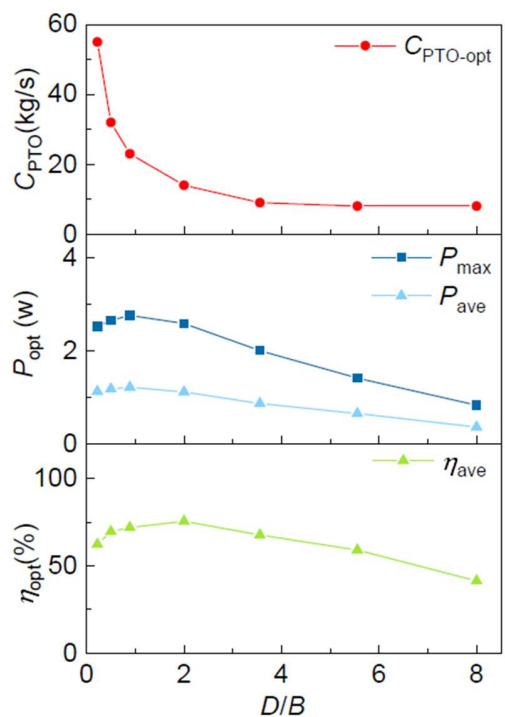

**Fig. 25.** Optimal PTO damping, maximum power capture and maximum conversion efficiency of S-WEC with buoys of draught-to-width ratios

Variation of the power generation and conversion efficiencies against the excitation frequency are further investigated for S-WECs with different buoy DWRs. A general damping coefficient $C_{PTO}$=20kg/s is taken for example. The excitation frequency varies from $0.66\omega'_1$ to $1.31\omega'_1$. From Fig. 26, it can be seen that as DWR increases from 0.22 to 0.89, the power captured by each buoy generally increases for all frequencies. The maximum $P_{ave}$ of a single buoy can reach 0.62W. From $D/B$=0.89 to 2.00, the maximum power generation does not vary evidently, but the power-frequency curve gets slimmer. As the DWR further decreases, the power captured by each buoy decreases over all the frequency range. Fig. 27 shows the conversion efficiencies corresponding to the power generation in Fig. 26. The efficiency-frequency and power-frequency curves have similar trends of deformation as DWR increases. With DWRs between 0.89 and 2.00, the buoys have the best performance on conversion efficiency in general, and the maximum conversion efficiency of each buoy can reach over 35%. It can be further confirmed that the buoy reaches the highest conversion efficiency when excited at its natural frequency.

*6.3. Effect of buoy's shape on power generation*

This subsection further discusses effects of the buoys' shape on their power generation capability. To be specific, four types buoys with the inclination angles of the bottom as $\alpha = 45°$, $60°$, $75°$ and $90°$, are considered for comparison, as illustrated in Fig. 28. The case of $\alpha = 90°$ is that adopted in physical experiments with $D/B = 0.5$. The width and displacement volume of these buoys are kept constant. Fundamental natural frequencies of the buoy and liquid in each S-WEC device are given in Fig. 29. It shows that, as the inclination angle increases from $\alpha = 45°$ to $90°$, the buoy's natural frequency is around $0.96\omega'$, which does not vary evidently. The natural frequency of the liquid in each S-WEC generally has a decrease trend with the increase of the inclination angle.

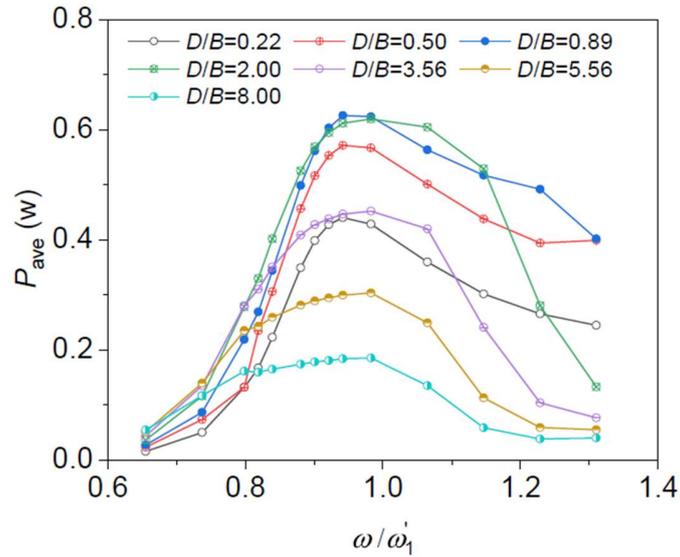

**Fig. 26.** Captured power of Buoy-1 with different draught-width-ratios, under different excitation frequencies

To figure out the optimal PTO damping of each buoy, the tank is excited at the buoy's natural frequency and the PTO damping is varied from $C_{PTO}$=5 to 100kg/s. Fig. 30 shows the captured power of Buoy-1 with different inclination angles of its bottom. It can be seen that for all these buoys an optimal PTO damping of $C_{PTO}$=30kg/s can lead to the largest power capture. As the inclination angle increases from $\alpha = 45°$ to $90°$, the buoy with $\alpha = 60°$ has the largest energy capture at all PTO damping.

Then, we fix the PTO damping at $C_{PTO}$=30kg/s and investigate the captured power and conversion efficiency of each buoy, when S-WEC is excited at various frequencies, referring to Figs. 31 and 32. Through comparison, it can be told that in the lower-frequency range with $\omega < 0.9\omega'_1$, both the power generation and conversion efficiency of each buoy are not affected evidently by the inclination angle. The captured power and conversion efficiency gradually reach the peak in a frequency range around $\omega = \omega\dot{\omega}'_1$. The pow-er or efficiency peak occurs at a higher excitation frequency for the buoy with a larger inclination bottom angle. As the excitation frequency further increases, the power generation tends to be steady with a slightly reducing magnitude, but the conversion efficiency drops evidently. Comparing these four cases, the buoy with the inclination angle of $\alpha = 60°$ has the best performance on the conversion efficiency in general, and the rectangular buoy (i.e.

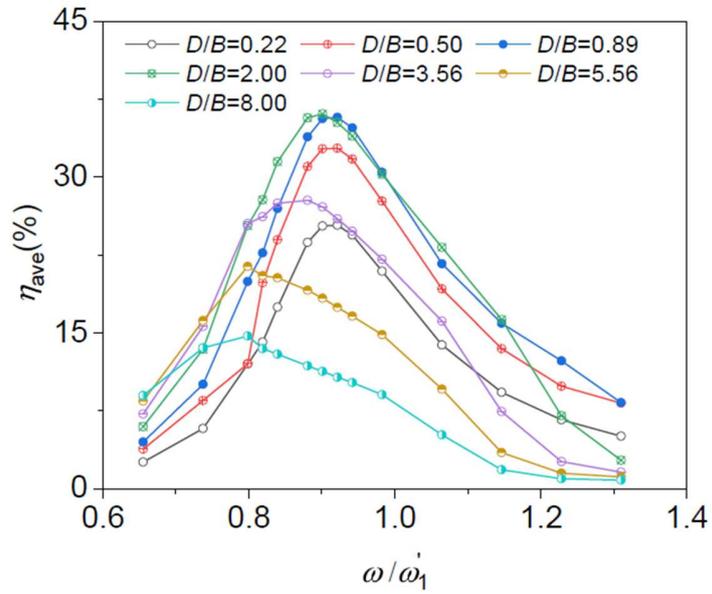

**Fig. 27.** Conversion efficiency of Buoy-1 with different draught-width-ratios, under different excitation frequencies

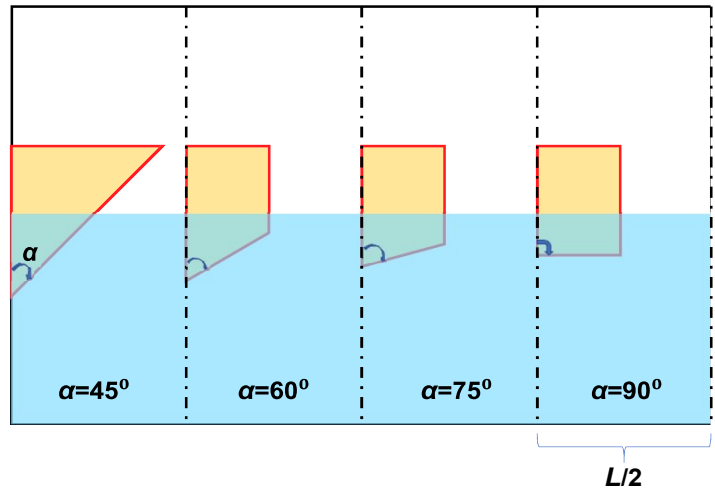

**Fig. 28.** Sketch of S-WECs with buoys of different bottom inclination

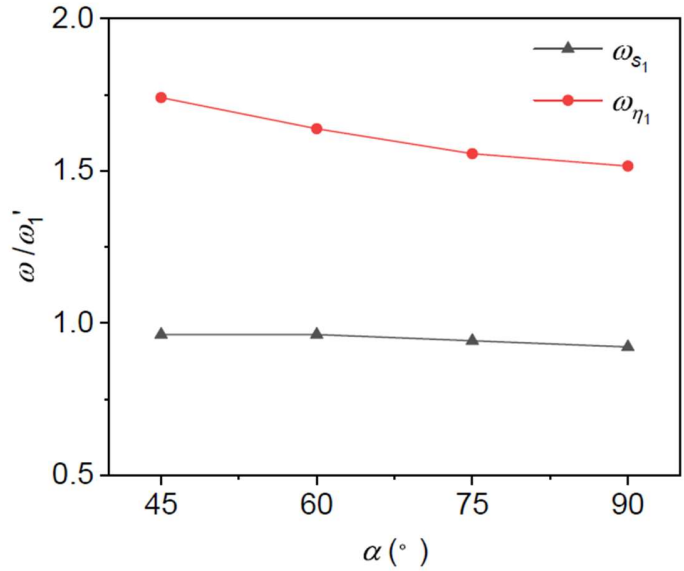

**Fig. 29.** Natural frequencies of the buoy and liquid in the S-WEC corresponding to different inclination angles of the buoy's bottom

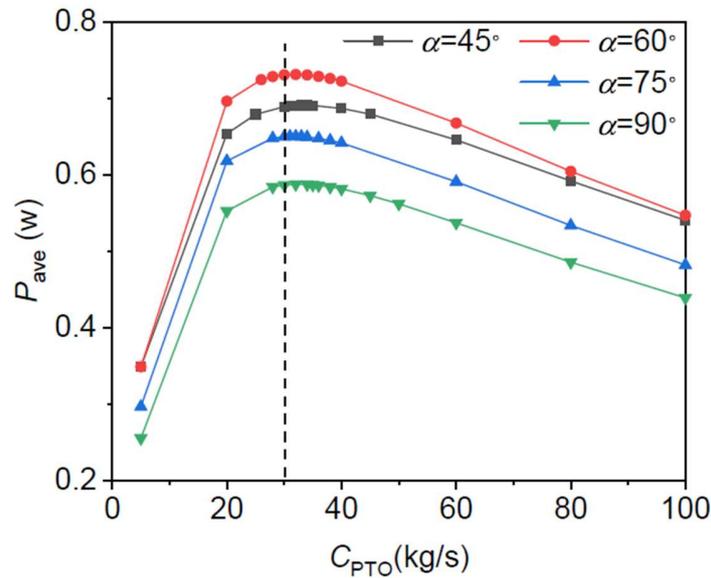

**Fig. 30.** Captured power of Buoy-1 with different inclination angles of the buoy's bottom, under different excitation frequencies

with α = 90°) does not show advantage on either the power generation or conversion efficiency. For S-WEC with two buoys, both the power generation and conversion efficiency can be double of the data in Figs. 31 and 32. To be specific, with α = 60°, over 3.6W can be generated in time average under the present small scale, and the conversion efficiency can reach over 90%. Such a capability of energy conversion is sufficiently high for wave energy converters.

## 7. Discussions on engineering practice

Besides those advantages mentioned in Section 1 on the enhancement of WEC's survivability and solution of technical compromise when integrating with floating structures, this section would like to give some further discus-sions on distinctive advantages of the present S-WEC design.

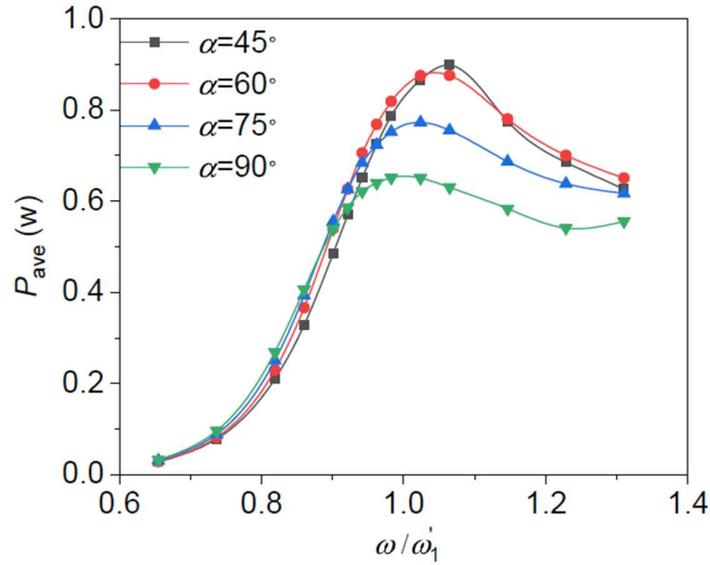

**Fig. 31.** Power generation of Buoy-1 with different inclination angles of the buoy's bottom

(1) The S-WEC is a fully-enclosed system, where the liquid tank can be half-filled with pure water or prepared solutions with certain density, and the air inside the tank can be Nitrogen, Carbon Dioxide or inert gases. With these applications, metal corrosion induced by oxidation and fouling organism inside the tank can be avoided.

(2) A liquid tank with clear internal surface is a natural energy storage. Take a conventional OB-WEC in open water for comparison. During an impulsive wave attack, only a limited proportion of wave energy can be captured and converted to an impulse in the electric signal, while the remaining wave energy is inevitably scattered and radiated away from the device without long-lasting retention. For the S-WEC, once initiated, sloshing waves in the tank can persist for a much longer duration, which means less wave energy can be less 'wasted' through wave scattering effects and the electric signal can have a better continuity.

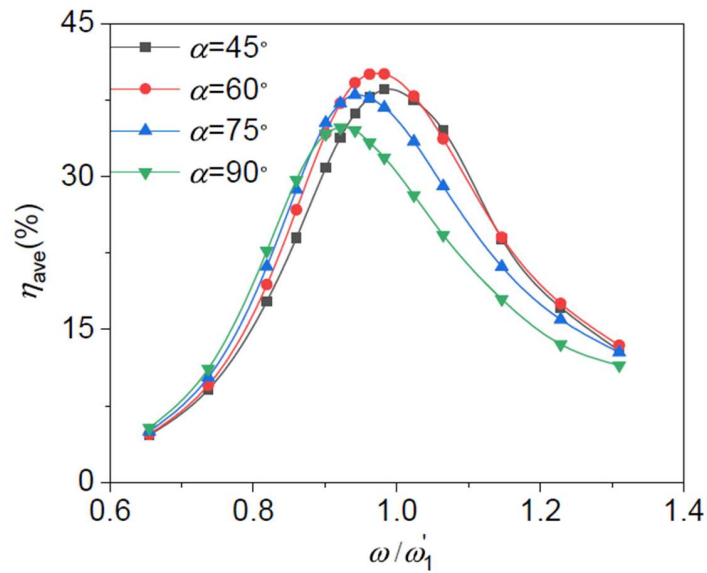

**Fig. 32.** Conversion efficiency of Buoy-1 with different inclination angles of the buoy's bottom

(3) The sloshing liquid is a natural frequency modulator of vibrations. Wave actions in real-sea states are full of randomness. Sloshing liquids canhelp filter out frequency components outside the predetermined dominant frequency spectra, leading to a smooth electricity signal.

(4) A liquid tank has infinite number of resonance frequencies. When the tank is exited around these frequencies, the external energy will continuously transfer to the tank liquid causing a violent sloshing for electricity generation. Thus, the S-WEC naturally has a wide range of frequency bandwidth for wave energy conversion. Unlikely, a conventional OB-WEC normally has one or finite number of best operation frequency, and people have to try every effort (through geometric modification or control strategies) to extend the available frequency bandwidth for energy conversion.

(5) Natural frequencies of the S-WEC system can be controlled easily by adjusting the liquid depth, so that it can adapt to a wide application scope of sea states.

## 8. Conclusions

A novel design of wave energy converter named 'S-WEC' is proposed to integrate with offshore floating structures. The mechanical system of S-WEC consists of a liquid tank with internal buoy-type WECs. When the floating structure is oscillated by external ocean waves, internal liquid sloshing is activated, and the mechanical energy of sloshing waves can be absorbed by the power take-off (PTO) system. A fully-nonlinear numerical model is established based on the boundary element method for a systematic investigation on dynamic properties of the proposed S-WEC. A motion decoupling algorithm based on auxiliary functions is developed to solve the nonlinear interaction of sloshing waves and floating buoys in the tank. Numerical convergence tests are carried out on the element size and time step for a stable simulation of the BEM solver. Physical experiments are carried out on a scaled S-WEC model to validate the mathematical and numerical methodology.

Natural frequencies of the S-WEC system are first investigated without consideration of damping-relevant factors, through spectrum analyses on motion histories of the buoy and sloshing liquid. The first two natural fundamental frequencies of the S-WEC system (i.e. $\omega_1$ and $\omega_2$) are identified, mainly relevant to motions of the buoys and liquids, respectively. Physically, if the S-WEC is excited at $\omega_1$, the buoys can oscillate in the resonant status, but the water-wave motion is very mild at the same time. This suggests the sloshing wave energy is largely transferred to the buoys in this condition, leaving aside a calm water domain. In the excitation condition of $\omega_2$, resonant liquid sloshing occurs accompanied by large-amplitude motion of buoys.

An artificial technique is applied in the potential-flow numerical method, in order to reflect the viscous damping effects of the sloshing liquid. Through comparisons with experimental measurements, the strength of the damping coefficient is calibrated. In the lower-frequency region of excitation, the motion amplitude increases sharply and shows a jump phenomenon towards the resonance condition. In the resonance condition, the motion amplitude of the buoy can reach about 6 times the amplitude of the supporting structure, which is an evident of the sloshing liquid on amplification of buoy's motion responses in waves. Liquid profiles of the S-WEC predicted by the present numerical method are compared with experimental snapshots at various time series, which confirms the reliability of present numerical solver on reproduction of the physical reality.

Effects of the PTO damping on power generation characteristics of S-WEC is further explored. The optimal PTO damping can be identified corresponding to the maximised power generation of buoys. It is found that different excitation frequencies correspond to different choices of the PTO damping. Within the considered frequency range of excitation, the optimal PTO damping first decreases and then increases, with the smallest value at the resonance condition. For each excitation frequency, the optimal PTO damping can lead to the maximisation of both the power generation and conversion efficiency of the buoy. If a constant PTO damping must be determined for the engineering design, this study provides a practical approach based on diagram analyses. With the selected value of PTO damping, the S-WEC can achieve at least 90% of the maximum power generation over the largest frequency range, corresponding the averaged conversion efficiency of 25% to 70%. In the higher-frequency regime, the conversion efficiency of the S-WEC can be around 50%.

Effects of the buoy's geometry on power generation characteristics of the S-WEC are also investigated. The geometry factor of draught-to-width ratios (DWRs) of each buoy is first considered. It shows that as DWR increases, both the fundamental natural frequencies and the optimal PTO damping of the S-WEC decrease. With the optimised PTO damping, buoys with the DWRs varying from 0.22 to 4.0 can lead to the conversion efficiency of 50 to 75%, and the best power generation performance appears at DWR=0.89. Then, the inclination angle $\alpha$ of the buoys bottom is considered, and effects of $\alpha$ on power generation capability of the buoys are investigated. As the inclination angle varies from 45° to 90°, neither the natural frequency nor the optimal PTO damping of the buoys vary greatly. For cases under consideration, the buoy with the inclination angle of 60° has the best performance on the conversion efficiency in general, and the conversion efficiency can reach over 90% which is sufficiently high for wave energy converters.

**Acknowledgement**

This study is supported by the National Key R & D Program of China (Grant No. 2018YFB1501904), National Natural Science Foundation of China (Grant No. 51709038) and the Project funded by China Postdoctoral Science Foundation (Grant No. 2019T120209).

**References**

[1] IPCC, Global warming of 1.5c. an ipcc special report on the impacts of global warming of 1.5c above pre-industrial levels and related global greenhouse gas emission pathways, in the context of strengthening the global response to the threat of climate change, sustainable development, and efforts to eradicate poverty [v. masson-delmotte, p. zhai, h. o. prtner, d. roberts, j. skea, p.r. shukla, a. pirani, w. moufouma-okia, c. pan, r.pidcock, s. connors, j. b. r. matthews, y. chen, x. zhou, m. i. gomis, e. lonnoy, t. maycock, m. tignor, t. waterfield (eds.)], In Press (2018).

[2] J. Rogelj, O. Geden, A. Cowie, A. Reisinger, Three ways to improve net-zero emissions targets, Nature 591 (2021) 365–368.

[3] J. Falnes, A review of wave-energy extraction, Marine Structures 20 (2007) 185 – 201.

[4] M. Mustapa, O. Yaakob, Y. M. Ahmed, C.-K. Rheem, K. Koh, F. A. Adnan, Wave energy device and breakwater integration: A review, Renewable and Sustainable Energy Reviews 77 (2017) 43 – 58.

[5] A. Pecher, J. Kofoed, Handbook of Ocean Wave Energy, volume 7, Springer, Germany, 2017.

[6] S. Astariz, G. Iglesias, The economics of wave energy: A review, Renewable and Sustainable Energy Reviews 45 (2015) 397 – 408.

[7] D. Vicinanza, P. Contestabile, J. Q. H. Nrgaard, T. L. Andersen, Innovative rubble mound breakwaters for overtopping wave energy conversion, Coastal Engineering 88 (2014) 154 – 170.

[8] C. Prez-Collazo, D. Greaves, G. Iglesias, A review of combined wave and offshore wind energy, Renewable and Sustainable Energy Reviews 42 (2015) 141 – 153.

[9] J. P. Kofoed, Wave Overtopping of Marine Structures: utilization of wave energy, Ph.D. thesis, Aalborg, 2002.

[10] P. Contestabile, C. Iuppa, E. D. Lauro, L. Cavallaro, T. L. Andersen, D. Vicinanza, Wave loadings acting on innovative rubble mound breakwater for overtopping wave energy conversion, Coastal Engineering 122 (2017) 60 – 74.

[11] S. Takahashi, H. Nakada, H. Ohneda, M. Shikamori, Wave power conversion by a prototype wave power extracting caisson in sakata port, in: Proceedings of the 23rd international conference on coastal engineering, 1992, pp. 3440–53.

[12] A. Brito-Melo, F. Neuman, A. Sarmento, Full-scale data assessment in OWC pico plant, Int J Offshore Polar Eng 18 (2008) 27 – 34.

[13] T. Whittaker, W. Beattie, M. Folley, C. Boake, A. Wright, M. Osterried, The LIMPET wave power project the first years of operation, Technical Report, 2004.

[14] J. R. Joubert, Design and development of a novel wave energy converter, Ph.D. thesis, Stellenbosch, 2013.

[15] I. Ortubia, L. Aguileta, J. Marqus, Mutriku wave power plant: from the thinking out to the reality, in: Proceedings of the 8th European wave and tidal energy conference, Uppsala, Sweden, 2009, pp. 319–29.

[16] F. Arena, A. Romolo, G. Malara, A. Ascanelli, On design and building of a U-OWC wave energy converter in the mediterranean sea: a case study, in: International Conference on Offshore Mechanics and Arctic Engineering, Volume 8: Ocean Renewable Energy, 2013, p. V008T09A102.

[17] D. Ning, R. Wang, Q. Zou, An experimental investigation of hydrodynamics of a fixed OWC wave energy converter, Appl Energy 168 (2016) 636 – 648.

[18] D. Howe, J.-R. Nader, OWC WEC integrated within a breakwater versus isolated: Experimental and numerical theoretical study, Interna-tional Journal of Marine Energy 20 (2017) 165–182.

[19] C. Xu, Z. Huang, A dual-functional wave-power plant for wave-energy extraction and shore protection: A wave-flume study, Applied Energy 229 (2018) 963–976.

[20] I. Lpez, B. Pereiras, F. Castro, G. Iglesias, Optimisation of turbine-induced damping for an OWC wave energy converter using a ransvof numerical model, Applied Energy 127 (2014) 105 – 114.

[21] R. Wang, D. Ning, C. Zhang, Q. Zou, Z. Liu, Nonlinear and viscous effects on the hydrodynamic performance of a fixed OWC wave energy converter, Coastal Engineering 131 (2018) 42 – 50.

[22] S. Mavrakos, G. Katsaounis, K. Nielsen, G. Lemonis, Numerical performance investigation of an array of heaving wave power converters in front of a vertical breakwater, in: Proceedings of the International Offshore and Polar Engineering Conference, 2004, pp. 238–245.

[23] J. Schay, J. Bhattacharjee, C. Guedes Soares, Numerical modelling of a heaving point absorber in front of a vertical wall, in: International Conference on Offshore Mechanics and Arctic Engineering, Volume 8: Ocean Renewable Energy, 2013, p. V008T09A0973.

[24] X. Zhao, D. Ning, D. Liang, Experimental investigation on hydrodynamic performance of a breakwater-integrated WEC system, Ocean Engineering 171 (2019) 25 – 32.

[25] C. Zhang, D. Ning, Hydrodynamic study of a novel breakwater with parabolic openings for wave energy harvest, Ocean Engineering 182 (2019) 540 – 551.

[26] H. Zhang, B. Zhou, C. Vogel, R. Willden, J. Zang, J. Geng, Hydrodynamic performance of a dual-floater hybrid system combining a floating

breakwater and an oscillating-buoy type wave energy converter, Applied Energy 259 (2020) 114212.

[27] C. Michailides, D. C. Angelides, Modeling of energy extraction and behavior of a flexible floating breakwater, Applied Ocean Research 35 (2012) 77 – 94.

[28] F. He, Z. Huang, A. W.-K. Law, An experimental study of a float- ing breakwater with asymmetric pneumatic chambers for wave energy extraction, Applied Energy 106 (2013) 222 – 231.

[29] A. Elhanafi, G. Macfarlane, A. Fleming, Z. Leong, Experimental and numerical investigations on the hydrodynamic performance of a floating-moored oscillating water column wave energy converter, Applied Energy 205 (2017) 369 – 390.

[30] A. Elhanafi, G. Macfarlane, D. Ning, Hydrodynamic performance of single-chamber and dual-chamber offshore-stationary oscillating water column devices using CFD, Applied Energy 228 (2018) 82 – 96.

[31] D. Ning, X. Zhao, M. Gteman, H. Kang, Hydrodynamic performance of a pile-restrained WEC-type floating breakwater: An experimental study, Renewable Energy 95 (2016) 531–541.

[32] X. Zhao, D. Ning, Experimental investigation of breakwater-type WEC composed of both stationary and floating pontoons, Energy 155 (2018) 226 – 233.

[33] F. Madhi, M. E. Sinclair, R. W. Yeung, The "berkeley wedge": an

asymmetrical energy-capturing floating breakwater of high performance, Marine Systems & Ocean Technology 9 (2014) 5–16.

[34] H. Zhang, B. Zhou, C. Vogel, R. Willden, J. Zang, L. Zhang, Hydrodynamic performance of a floating breakwater as an oscillating-buoy type wave energy converter, Applied Energy 257 (2020) 113996.

[35] N. Fonseca, J. Pessoa, Numerical modeling of a wave energy converter based on U-shaped interior oscillating water column, Applied Ocean Research 40 (2013) 60 – 73.

[36] S. R. e Silva, R. Gomes, A. Falco, Hydrodynamic optimization of the ugen: Wave energy converter with u-shaped interior oscillating water column, International Journal of Marine Energy 15 (2016) 112 – 126.

[37] S. Crowley, R. Porter, D. Evans, A submerged cylinder wave energy converter with internal sloshing power take off, European Journal of Mechanics - B/Fluids 47 (2014) 108 – 123.

[38] N. M. Tom, F. Madhi, R. W. Yeung, Power-to-load balancing for heav- ing asymmetric wave-energy converters with nonideal power take-off, Renewable Energy 131 (2019) 1208 – 1225.

[39] C. Zhang, J. Tan, D. Ning, B. Liang, A wave energy conversion device based on liquid tanks with internal floating buoys for integratation with floating breakwater (China Patent, CN201821636196.3, June 2019).

[40] M. E. McCormick, Ocean Wave Energy Conversion, Dover Publications, New York, 2007.

[41] D. Sen, Numerical simulation of motions of two-dimensional floating bodies, Journal of Ship Research 37 (1993) 307–330.

[42] G. Wu, R. Eatock Taylor, The coupled finite element and boundary element analysis of nonlinear interactions between waves and bodies, Ocean Engineering 30 (2003) 387–400.

[43] J. B. Frandsen, Sloshing motions in excited tanks, Journal of Computational Physics 196 (2004) 53 – 87.

[44] C. Zhang, Nonlinear simulation of resonant sloshing in wedged tanks using boundary element method, Engineering Analysis with Boundary Elements 69 (2016) 1 – 20.